\documentstyle[aps,amssymb,epsfig,preprint,eqsecnum,floats,cite]{revtex}
\tighten
\def \be{\begin{equation}}
\def \ee{\end{equation}}
\def \bea{\begin{eqnarray}}
\def \eea{\end{eqnarray}}
\def \ben{\begin{enumerate}}
\def \een{\end{enumerate}}
\def \bit{\begin{itemize}}
\def \eit{\end{itemize}}
\def \baR{\begin{array}}
\def \eaR{\end{array}}
\def \Box{\text{box}}
\def \Count{\text{count}}
\def \Peng{\text{peng}}
\def \chargino{\tilde{\chi}^{\pm}}
\def \sneutrino{\tilde{\n}}
\def \tsquark{\tilde{t}}
\def \upsquark{\tilde{u}}
\def \av#1{\left\langle #1\right\rangle}

\def \B{\bar{B}}
\def \branch{{\mathcal B}\,}
\def \cl#1{{#1\%\ \mathrm{C.L.}}}
\def \heff{H_{\text{eff}}}

\def \Re{{\text{Re}}\,}

\def \GeV{{\text{GeV}}}

\def \MeV{{\text{MeV}}}

\def \bm{\boldmath}

\def \braket#1#2#3{\langle #1|#2| #3\rangle}

\def \cp{\mathrm{CP}}

\def \diag{{\mathrm{diag}}}
\def \dis{\displaystyle}
\def \ea{{\it et al.}}

\def \eq#1{Eq.~(\ref{#1})}
\def \eqs#1#2{Eqs.~(\ref{#1})--(\ref{#2})}
\def \fig#1{Fig.~\ref{#1}}

\def \heff{H_{\text{eff}}}
\def \nnu{\nonumber}

\def \Oi{{\mathcal O}}
\def \ol#1{\overline{#1}}

\def \sm{\mathrm{SM}}
\def \rf{Ref.~\cite}
\def \rfs{Refs.~\cite}
\def \sec#1{Sec.~\ref{#1}}
\def \Vec#1{{\bf#1}}
\def \cseff {c_7^{\text{eff}}}
\def \ceff {c_9^{\text{eff}}}

\def \cten{c_{10}}
\def \a{\alpha}
\def \b{\beta}

\def \g{\gamma}
\def \G{\Gamma}

\def \la{\lambda}

\def \m{\mu}
\def \n{\nu}

\def \p{\pi}
\def \r{\rho}
\def \s{\sigma}

\def \t{\tau}
\def \pslash{p\hspace{-0.45em} /}

\def\euro#1#2#3{{Eur. Phys. J. C} {\bf #1}, #3 (#2)}

\def\ibid#1#2#3{{\it ibid.\/}~{\bf#1}, #3 (#2)}
\def\ib#1#2#3{{\bf#1}, #3 (#2)}
\def\jphys#1#2#3{{J.~Phys. G} {\bf #1}, #3 (#2)}

\def\nim#1#2#3{{Nucl.~Instrum.~Methods Phys.~Res.}~{\bf A#1}, #3 (#2)}
\def\np#1#2#3{{Nucl.~Phys.}~{\bf B#1}, #3 (#2)}

\def\pl#1#2#3{{Phys.~Lett. B}~{\bf #1}, #3 (#2)}

\def\prd#1#2#3{{Phys.~Rev. D}~{\bf #1}, #3 (#2)}
\def\prl#1#2#3{{Phys.~Rev.~Lett.}~{\bf #1}, #3 (#2)}
\def\prp#1#2#3{{Phys.~Rep.}~{\bf #1}, #3 (#2)}
\def\ptp#1#2#3{{Prog. Theor.~Phys.}~{\bf #1}, #3 (#2)}
\def\rmp#1#2#3{{Rev. Mod. Phys.} {\bf #1}, #3 (#2)}
\def\zpc#1#2#3{{Z.~Phys. C}~{\bf #1}, #3 (#2)}
\begin{document}
\preprint{\setlength{\baselineskip}{1.5em}
%\small
\vbox{\vspace{-1cm}
\hbox{TUM-HEP-411/01}
\hbox{hep-ph/0104284}
\hbox{April 2001}}}
\draft
\title{Analysis of Neutral Higgs-Boson Contributions to the Decays \\
\bm$\bar{B}_s \to l^+l^-$ and \bm$\bar{B}\to K l^+l^-$}
\author{\sc C.~Bobeth\thanks{E-mail address: bobeth@ph.tum.de}, 
T.~Ewerth\thanks{E-mail address: tewerth@ph.tum.de}, 
F.~Kr\"uger\thanks{E-mail address: fkrueger@ph.tum.de}, and 
J.~Urban\thanks{E-mail address: urban@ph.tum.de}}
\address{Physik Department, Technische Universit\"at M\"unchen, 
D-85748 Garching, Germany}
%
%\date{\today}
%
\maketitle
\begin{abstract}
We report on a calculation of Higgs-boson contributions to 
the decays $\bar{B}_s \to l^+l^-$ and $\bar{B}\to K l^+l^-$ 
$(l=e,\m)$ which are governed by the effective Hamiltonian describing 
$b\to s l^+ l^-$.~Compact formulae for the Wilson  
coefficients are provided  
in the context of the type-II 
two-Higgs-doublet model 
(2HDM) and supersymmetry (SUSY) with minimal flavour violation, focusing 
on the case of large $\tan\b$.~We derive, 
in a model-independent way, constraints on
Higgs-boson-mediated interactions, using present experimental results on rare 
$B$ decays including $b\to s \g$, $\bar{B}_s \to \m^+\m^-$, and 
$\B\to K^{(*)}\m^+\m^-$. In particular, we assess the 
impact of possible scalar and pseudoscalar interactions 
transcending the standard model (SM)  
on the branching ratio of $\bar{B}_s \to \m^+\m^-$ and the  
forward-backward (FB) asymmetry of $\m^-$ in 
$\bar{B}\to K \m^+\m^-$ decay. The average 
FB asymmetry, which is unobservably small within the SM, and therefore 
a potentially valuable tool to search for new physics,  
is predicted to be no greater than 
$4\%$ for a nominal branching ratio of about $6\times 10^{-7}$.
Moreover, striking effects on the decay spectrum of $\B\to K\m^+\m^-$ are 
already ruled out by experimental data on the $\bar{B}_s \to \m^+\m^-$ branching fraction.
In addition, we study the constraints on the parameter space of the 
2HDM and SUSY with minimal flavour violation. While the type-II 2HDM
does not give any sizable contributions to the above decay modes,  
we find that SUSY contributions obeying the constraint 
on $b\to s \g$ can significantly affect  the branching 
ratio of $\bar{B}_s \to \m^+\m^-$. We also comment on previous 
calculations contained in the literature.
\end{abstract}
%
%%%%%%%%%%%%%% END OF TITLE PAGE %%%%%%%%%%%%%%%%%%%%%
%
\section{Introduction}
At the quark level, the decays $\B_s\to l^+l^-$ and $\B\to K l^+ l^-$, where
$l$ denotes either $e$ or $\m$, are generated by the short-distance 
effective Hamiltonian for $b\to s l^+l^-$.\footnote{We adopt a convention where $\B\equiv (\bar{d}b)$ or $(\bar{u}b)$ and $\B_s\equiv (\bar{s}b  )$.} 
Within the standard model (SM), 
the decay $\B_s\to l^+l^-$ proceeds via $Z^0$ penguin and box-type 
diagrams, and its branching ratio is expected to be highly suppressed. 
Likewise, the forward-backward (FB) asymmetry of the 
lepton in $\B\to K l^+ l^-$ is exceedingly small. 
However, in models with an extended Higgs sector these observables may 
receive sizable contributions, and thus provide 
a good opportunity to look for new physics.
The models to be considered are a type-II two-Higgs-doublet model (2HDM) and a 
supersymmetric extension of the SM with minimal flavour violation 
(see, e.g., \rfs{mssm:original,higgs:hunters}) -- that is, 
we assume the Cabib\-bo-Ko\-ba\-ya\-shi-Maskawa (CKM) matrix to be the only source of flavour mixing.
An interesting feature of these models is that large values of
$\tan\b$, the ratio of the two vacuum expectation values of the 
neutral Higgs fields, may compensate for the inevitable suppression by 
the mass of the light leptons $e$ or $\m$.

The calculation of  Higgs-boson exchange diagrams contributing to the  
$b\to s l^+l^-$ transition has been the subject of many investigations
\cite{bll:SM:buch:buras,bll:2hdm,skiba:kali,grossman:etal,large:tan:beta:bll,guetta:nardi,logan:nierste,huang:etal:recent,chan:slaw}. As pointed out in  
\rf{logan:nierste}, the results obtained in the context of the 2HDM 
disagree with each other.
In view of this, we re-analyse the 
$b\to s l^+l^-$ transition, confining ourselves to the case of large $\tan\b$
in the range $40\leqslant \tan\b\leqslant 60$. Our study extends 
previous analyses in several ways. We include, for example, other 
rare $B$ decays in addition to $b\to s \g$ 
to constrain possible scalar and pseudoscalar 
interactions outside the SM. We also assess their contributions 
to various observables in $b\to sl^+ l^-$ transitions  such as 
$\B\to K^{(*)}l^+ l^-$.
     
So far all experimental results yield only upper bounds on the 
decay modes governed by $b\to sl^+ l^-$. 
The best upper limits at present come 
from processes with muons in the final state, and we therefore 
concentrate on the $\m^+\m^-$ mode.
Specifically, we address the viability of the short-distance 
coefficients in the presence of scalar and pseudoscalar interactions
with the measured $b\to s \g$ rate and the 
experimental bound $\branch(\B_s\to \m^+\m^-)<2.6\times 
10^{-6}$ ($\cl{95}$) \cite{exp:bstomumu}, as well as with the restrictions 
imposed by the upper limits on $\B\to K^{(*)}\m^+\m^-$ 
\cite{limits:exc:exp}.

The outline of the paper is as follows. The effective Hamiltonian 
des\-cri\-bing the quark transition $b\to s l^+l^-$ in the presence of 
non-standard Higgs bosons is reviewed in \sec{eff:ham}.
In \sec{had:mat}, we discuss the hadronic matrix elements required for 
the decays 
$\B_s\to l^+l^-$ and $\B\to K l^+l^-$. The corresponding angular 
distributions and decay spectra are presented in \sec{dec:dis}. 
Section \ref{computation} is devoted to the calculation of 
the Higgs-boson diagrams in a general $R_\xi$ gauge, and also contains 
a brief description of 
our renormalization procedure. Readers who are not particularly 
interested in the details of the computation, can skip this part and 
proceed to the discussion of our results, which are obtained in the framework 
of the type-II 2HDM and supersymmetry (SUSY) 
with minimal flavour violation.
Attention is focused on the interesting case of large $\tan\b$, 
and a comparison is made with the results of previous studies.
In \sec{num:analysis}, we derive model-independent upper bounds on 
scalar and pseudoscalar interactions, and explore their implications for the 
branching fraction of $\B\to K \m^+\m^-$, as well as the corresponding FB 
asymmetry, using presently available data on the decays  
$b\to s\g$, $\B\to K^{(*)} \m^+\m^-$, and $\B_s\to \m^+\m^-$. As an 
application, we investigate the constraints on the parameter space in the 
aforementioned extensions of the SM. 
Finally, in \sec{discussion}, we present some concluding remarks. 
%
%An appendix lists explicit formulae for the Wilson coefficients which 
%are valid at large $\tan\beta$.
%

\section{Effective Hamiltonian for \bm \lowercase{$b\to sl^+l^-$}}\label{eff:ham}
The starting point of our analysis is the effective Hamiltonian describing 
$b\to s \l^+l^-$: 
\bea\label{heff}
\heff = -\frac{4G_F}{\sqrt{2}}V_{tb}^{}V_{ts}^{*}
\Bigg\{
&\dis\sum\limits_{i = 1}^{10}&c_i (\m)\Oi_i(\m)+ c_S(\m) \Oi_S(\m) + 
c_P(\m) \Oi_P(\m)\nnu\\
&+&c_S'(\m) \Oi_S'(\m) + 
c_P'(\m) \Oi_P'(\m)\Bigg\},
\eea
where $c_i^{(\prime)}(\m)$ and $\Oi_i^{(\prime)}(\m)$ 
are the Wilson coefficients and 
local operators respectively. In writing \eq{heff}, we have used the 
unitarity of the CKM matrix and omitted terms proportional to 
$V_{ub}^{}V^*_{us}/V_{tb}^{}V^*_{ts}\sim O(10^{-2})$.
Taking the limit $c_{S,P}^{(\prime)}\to 0$, 
we recover the effective Hamiltonian of the SM 
\cite{eff:ham:sm,review}.

The evolution of the short-distance 
coefficients evaluated at the matching scale $\m_W=M_W$ down to the 
low-energy scale at $\m_b=m_b^{\text{pole}}$   can be performed
using renormalization group equations
(see, e.g., \rf{review}).
The operator basis is given by
\bea\label{operatorbasis}
{\Oi_1}&=& (\bar{s}_{ \alpha} \gamma_\mu P_L c_{ \beta})
(\bar{c}_{ \beta} \gamma^\mu P_Lb_{ \alpha}),\nnu\\
{\Oi}_2 &=& (\bar{s}_{ \alpha} \gamma_\mu P_L c_{ \alpha})
(\bar{c}_{ \beta} \gamma^\mu P_L b_{ \beta}),\nnu\\
{\Oi}_3 &=& (\bar{s}_{ \alpha} \gamma_\mu P_L b_{ \alpha})\sum_{q=u,d,s,c,b}
(\bar{q}_{ \beta} \gamma^\mu P_L q_{ \beta}),\nnu\\
{\Oi}_4&=& (\bar{s}_{ \alpha} \gamma_\mu P_L b_{ \beta})
\sum_{q=u,d,s,c,b}(\bar{q}_{ \beta} \gamma^\mu P_L q_{ \alpha}),\nnu\\
{\Oi}_5&=& (\bar{s}_{ \alpha} \gamma_\mu P_L b_{ \alpha})
\sum_{q=u,d,s,c,b}(\bar{q}_{ \beta} \gamma^\mu P_R q_{ \beta}),\nnu\\
{\Oi}_6&=& (\bar{s}_{ \alpha} \gamma_\mu P_L b_{ \beta})
\sum_{q=u,d,s,c,b}(\bar{q}_{ \beta} \gamma^\mu P_R q_{ \alpha}),\nnu\\ 
{\Oi}_7&=&\frac{e}{16 \pi^2}m_b (\bar{s}_{\alpha} \sigma_{\mu \nu}
P_R b_{\alpha})F^{\mu \nu},\nnu\\
{\Oi}_8& =& \frac{g_s}{16 \pi^2}m_b (\bar{s}_{\alpha} 
T_{\alpha \beta}^a \sigma_{\mu \nu}P_R b_{\beta}) G^{a \mu \nu},\nnu\\
{\Oi}_9&=& \frac{e^2}{16 \pi^2} (\bar{s}_\alpha \gamma^{\mu} P_L b_\alpha)
(\bar{l} \gamma_{\mu} l),\nnu\\
{\Oi}_{10}&=&\frac{e^2}{16 \pi^2} (\bar{s}_\alpha \gamma^{\mu} P_L
b_\alpha) (\bar{l} \gamma_{\mu}\gamma_5 l),\nnu\\
{\Oi}_S&=&\frac{e^2}{16 \pi^2} m_b (\bar{s}_\alpha P_R b_\alpha) 
(\bar{l}l),\nnu\\
{\Oi}_P&=&\frac{e^2}{16 \pi^2} m_b (\bar{s}_\alpha P_R b_\alpha)(\bar{l} 
\gamma_5 l),\nnu\\
{\Oi}'_S&=&\frac{e^2}{16 \pi^2} m_s (\bar{s}_\alpha P_L b_\alpha) 
(\bar{l}l),\nnu\\
{\Oi}'_P&=&\frac{e^2}{16 \pi^2} m_s (\bar{s}_\alpha P_L b_\alpha)
(\bar{l} \gamma_5 l),
\eea
$\a$, $\b$ being colour indices, $a$ labels  the SU(3) generators,
and $P_{L,R}= (1\mp \g_5)/2$. Notice that we have dropped the $m_s$ 
corrections to $\Oi_7$ and $\Oi_8$ 
while retaining the primed operators
 $\Oi_S'$ and $\Oi_P'$.
Further discussion on this point will be given in \sec{computation}.
In general, there are additional operators such as
$(\bar{s} \s_{\m \n}P_{L,R} b)(\bar{l}\s^{\m \n}P_{L,R}l)$ which, as we will argue later, 
do not contribute to the decay $\B_s\to l^+l^-$ 
but show up in the process $\B\to K l^+l^-$. 
However,  these operators are not expected to contribute significantly
\cite{int:tensor}, and we shall neglect them in our subsequent discussion.  

\section{Hadronic matrix elements}\label{had:mat}
\subsection{\bm $\B\to K l^+l^-$}
The hadronic matrix elements responsible for the exclusive 
decay $\B\to K l^+l^-$ are conveniently defined as 
\cite{stech:wirbel,ali:etal,melikhov:stech}
\be\label{matrix:element:btok:I}
\braket{K(k)}{\bar{s}\g_{\m} b}{\B(p)}=
(2p-q)_{\m} f_+(q^2)+ \frac{M_B^2-M_K^2}{q^2}q_{\m}[f_0(q^2)-
f_+(q^2)], 
\ee
\be\label{matrix:element:btok:II}
\braket{K(k)}{\bar{s}i\s_{\m\n}q^{\n}b}{\B(p)}=
- [(2p - q)_{\m}q^2 - (M_B^2 - M_K^2)q_{\m}]\frac{f_T(q^2)}{M_B+M_K},
\ee
where $q^\m=(p-k)^\m$ is the four-momentum transferred to the 
dilepton system.\footnote{Note that
$\braket{K}{\bar{s}\g_{\m}(1-\g_5) b}{\B}=
\braket{K}{\bar{s}\g_{\m}(1+\g_5) b}{\B}
\equiv \braket{K}{\bar{s}\g_{\m} b}{\B}$.} 
Further, employing the equation of motion for $s$ and $b$ quarks, we obtain, 
from \eq{matrix:element:btok:I},  
\be\label{matrix:element:btok:III}
\braket{K(k)}{\bar{s} b}{\B(p)}=\frac{M_B^2-M_K^2}{m_b-m_s}f_0(q^2).
\ee
In the following we shall adopt the form factors of 
\rf{ali:etal}, which uses light cone sum rule results.
 
\subsection{\bm$\B_s\to l^+l^-$}
The relevant matrix elements are characterized by the decay constant of the 
pseudoscalar meson $\B_s$, which is defined by the axial vector 
current matrix element
\cite{skiba:kali,grossman:etal,guetta:nardi}:
\be\label{matrix:element:btoll:I}
\braket{0}{\bar{s}\g_{\m}\g_5 b}{\B_s(p)}=ip_{\mu}f_{B_s},
\ee
while the  matrix element of the vector current vanishes. (For the decay 
constant $f_{B_s}$ we will take the value $210\pm 30\ \MeV$ \cite{fbs}.)
Contracting both sides in \eq{matrix:element:btoll:I} with $p^\m$ and 
using the equation of motion gives
\be\label{matrix:element:btoll:II}
\braket{0}{\bar{s}\g_5 b}{\B_s(p)}=- i f_{B_s} \frac{M_{B_s}^2}{m_b+m_s}.
\ee
An important point to note is that the matrix element in 
\eq{matrix:element:btoll:I} vanishes when contracted with 
the leptonic vector current $\bar{l}\g_\m l$ as it is proportional to 
$p^\m= p^\m_{l^+} + p^\m_{l^-}$, which is the only vector that can be 
constructed. In addition,
the matrix element $\braket{0}{\bar{s}\s_{\m\n} b}{\B_s(p)}$ must vanish 
since it is not possible to construct a combination made up of $p^\m$
that is antisymmetric with respect to the index 
interchange $\m\leftrightarrow\n$.
Consequently, the operators $\Oi_7$ and $\Oi_9$ do not contribute to 
the decay $\B_s\to l^+l^-$, which is then governed by
$\Oi_{10}$ and $\Oi_{S,P}^{(\prime)}$ defined in \eq{operatorbasis}.

\section{Differential decay distributions}\label{dec:dis}
Using \eq{heff} together with  
\eqs{matrix:element:btok:I}{matrix:element:btoll:II}, 
the matrix element for the just-mentioned 
decay modes can be written in the form 
\be
{\mathcal M}=\frac{ G_F \a}{\sqrt{2}\p}V_{tb}^{}V_{ts}^{\ast}
\Bigg[F_S \bar{l}l + F_P \bar{l}\g_5 l  + 
F_V p^{\m} \bar{l}\g_{\m}l+ F_A p^{\m} \bar{l}\g_{\m}\g_5 l\Bigg],
\ee 
$p^\m$ being the four-momentum of the initial $B$ meson,
and the $F_i$'s are functions of Lorentz-invariant quantities. 
It should be emphasized that the form factors $F_S$ and $F_P$ must 
vanish when $m_l = 0$ because of chiral symmetry, 
hence $F_{S,P}\propto m_l$.~Nevertheless, as will be elaborated below, 
large values of $\tan\b$ may compensate for the suppression by the
electron or muon mass in certain extensions of the SM.

Squaring the matrix element and summing over lepton spins, we find the result
\bea\label{mes:general}
|{\mathcal M}|^2 &\propto & (s-4m_l^2) |F_S|^2 + s  |F_P|^2 + [4 (p\cdot p_{l^-}) 
(p\cdot p_{l^+})-M^2 s] ( |F_A|^2+ |F_V|^2)+ 4m_l^2 M^2 |F_A|^2\nnu\\
&+& 4m_l [p\cdot(p_{l^+}- p_{l^-})\Re (F_S F_V^*) + (p\cdot q) \Re(F_P F_A^*)],
\eea
where $s\equiv q^2$, $q=p_{l^+}+p_{l^-}$, 
and $M$ refers to the mass of the decaying $B$ meson.

\subsection{\bm $\B\to K l^+l^-$}
Let us start with the decay $\B\to K l^+l^-$, where we will employ 
the definitions
\be
\G_0=\frac{G_F^2 \a^2}{2^{9} \p^5 M_B^3} |V_{tb}^{}V_{ts}^{\ast}|^2,
\ee
\be
\la (a,b,c) = a^2 + b^2 + c^2 - 2 (a b + b c + a c), \quad 
\b_l=\sqrt{1-4 m_l^2/s}.
\ee
Furthermore, we define $\theta$ as the angle  between the three-momentum 
vectors $\Vec{p}_{l^-}$ and $\Vec{p}_s$ in the dilepton centre-of-mass 
system. 
The two-dimensional spectrum is then given by 
\bea\label{dist:two-fold}
&&\frac{1}{\G_0}\frac{d\G(\B\to K l^+ l^-)}{d s\, d\!\cos\theta}
=\la^{1/2}(M_B^2,M_K^2,s)\b_l \Bigg\{s(\b_l^2 |F_S|^2 +  |F_P|^2) + 
\frac{1}{4}\la(M_B^2,M_K^2,s)\nnu\\ 
&&\quad\mbox{} \times [1-\b_l^2 \cos^2\theta] (|F_A|^2+ |F_V|^2)
+ 4m_l^2 M_B^2 |F_A|^2+ 2m_l [\la^{1/2}(M_B^2,M_K^2,s)\b_l 
\Re (F_S F_V^*)\cos\theta \nnu\\ 
&&\quad \mbox{}+ (M_B^2-M_K^2+s)\Re(F_P F_A^*)]\Bigg\},
\eea
with $s$ and $\cos\theta$ bounded by
\be 
4 m_l^2 \leqslant s \leqslant (M_B-M_K)^2, \quad
-1 \leqslant \cos\theta\leqslant 1.
\ee

A quantity of particular interest is the forward-backward asymmetry
\be\label{FB}  
A_{\text{FB}}(s)=\frac{\dis\int_0^1 d\!\cos\theta
\frac{d\G}{d s\, d\!\cos\theta}-\int_{-1}^0 d\!\cos\theta
\frac{d\G}{d s\, d\!\cos\theta}}{\dis\int_0^1 d\!\cos\theta
\frac{d\G}{d s\, d\!\cos\theta}+\int_{-1}^0 d\!\cos\theta
\frac{d\G}{d s\, d\!\cos\theta}},
\ee
which is given by 
\be\label{FB:explicitly} 
A_{\text{FB}}( s)=\frac{2 m_l \la (M_B^2,M_K^2,s) 
\b_l^2 \Re(F_S F_V^*) \G_0}{d\G/ ds },
\ee
where the dilepton invariant mass spectrum, $d\G/ds$, can be obtained 
by integrating the distribution in \eq{dist:two-fold} with respect to 
$\cos\theta$. Explicitly, we find 
\bea
&&\frac{1}{\G_0}\frac{d\G(\B\to K l^+ l^-)}{d s}
=2 \la^{1/2}(M_B^2,M_K^2,s)\b_l
\Bigg\{s(\b_l^2 |F_S|^2 +  |F_P|^2) + 
\frac{1}{6}\la(M_B^2,M_K^2,s)\nnu\\
&&\quad\times \Bigg(1 + \frac{2m_l^2}{s}\Bigg)
(|F_A|^2+ |F_V|^2)+ 4m_l^2 M_B^2 |F_A|^2+ 2m_l(M_B^2-M_K^2+s)\Re(F_P F_A^*)]
\Bigg\}.
\eea

The Lorentz-invariant functions $F_i$ in the above formulae  
depend on the Wilson coefficients as well as the $s$-dependent form factors 
introduced in the preceding sections,  namely,
\be\label{fs:btokmumu}
F_S= \frac{1}{2}(M_B^2-M_K^2)f_0(s)
\Bigg[\frac{c_S m_b + c'_S m_s}{m_b-m_s}\Bigg],
\ee
\be
F_P=-m_l \cten\Bigg\{f_+(s) - \frac{M_B^2-M_K^2}{s}[f_0(s)-f_+(s)]\Bigg\}
+ \frac{1}{2}(M_B^2-M_K^2)f_0(s)\Bigg[\frac{c_P m_b+c'_P m_s}{m_b-m_s}\Bigg],
\ee
\be\label{fa:btokmumu}
F_A= \cten f_+(s), \quad F_V= \Bigg[\ceff f_+(s) + 2 \cseff m_b 
\frac{f_T(s)}{M_B+M_K}\Bigg].
\ee
It should be noted that within the SM the Wilson coefficients of scalar and 
pseudoscalar operators are invariably suppressed by $m_l m_{b,s}/M_W^2$, 
leading to $c_{S,P}^{(\prime)}\simeq 0$, and so the FB asymmetry vanishes.
This observable is therefore particularly useful for testing non-SM physics
and its measurement could  provide vital information on 
an extended Higgs sector. 

The analytic expressions for the remaining coefficients appearing in 
\eqs{fs:btokmumu}{fa:btokmumu} may be found in \rfs{eff:ham:sm,review}. Within the SM, they are estimated to 
be\footnote{We use a running 
top-quark mass of $m_t\equiv \ol{m}_t(m_t)=166\pm 5\ \GeV $, corresponding to 
$m^{\text{pole}}_t=174.3\pm 5.1\ \GeV$ \cite{PDG}.}
\be\label{coeffs:sm} 
\cseff=-0.310, \quad \ceff = c_9 + Y(s), \quad c_9= 4.138, \quad \cten=-4.221,
\ee
where the function $Y(s)$ denotes the contributions from the one-loop 
matrix elements of the four-quark operators $\Oi_1$--$\Oi_6$
(see Appendix \ref{aux:funcs}). 

Finally, we give here the SM prediction of the non-resonant 
branching fraction for the decay into  a $\m^+\m^-$ pair, the result being  
\be
\branch(\B\to K \m^+\m^-)= (5.8\pm 1.2)\times 10^{-7},
\ee
where the error is due to the uncertainty in the hadronic form factors,
which is the major source of uncertainty in the branching 
ratio. We do not 
address here the issue of resonances such as $J/\psi, \psi'$, which 
originate from real $c\bar{c}$ intermediate states. 
For theoretical discussions of these contributions and the various 
approaches proposed in the literature, the reader is referred 
to \rfs{ali:etal,long-distance}.

\subsection{\bm$\B_s\to l^+ l^-$}
Our results for the matrix element squared [\eq{mes:general}] are 
immediately adaptable to the process $\B_s\to l^+ l^-$. Using
$p= p_{l^+} + p_{l^-}$, we obtain the branching ratio
\be\label{BR:bll}
\branch(\B_s\to l^+ l^-)=\frac{G_F^2 \a^2 M_{B_s}\t_{B_s}}{16 \p^3}
|V_{tb}^{}V_{ts}^{\ast}|^2 \sqrt{1-\frac{4m_l^2}{M_{B_s}^2}}\Bigg\{
\Bigg(1-\frac{4m_l^2}{M_{B_s}^2}\Bigg)|F_S|^2 +
|F_P + 2 m_l F_A|^2\Bigg\}.
\ee
The factor $m_l$ in front of $F_A$ reflects the fact that 
within the SM the decays $\B_s\to e^+ e^-$ or $\m^+\m^-$ are helicity 
suppressed due to angular momentum conservation; indeed, since the 
$B$ meson is spinless, both $l^+$ and $l^-$ must 
have the same helicity.

The scalar, pseudoscalar, and axial vector form factors are given by 
$(i=S,P)$
\be\label{fi:btomumu}
F_i=-\frac{i}{2}M_{B_s}^2 f_{B_s}\Bigg[\frac{c_im_b-c_i' m_s}{m_b+m_s}\Bigg],
\quad F_A= - \frac{i}{2}f_{B_s}\cten .
\ee
Throughout the present paper we use the leading-order 
result for the Wilson coefficient $\cten$ in order to be consistent with the 
precision of the calculation that will be presented in \sec{computation}. 
This is different from \rfs{logan:nierste,chan:slaw} 
where the next-to-leading-order result for the SM contribution
 has been used.

For completeness, let us record the SM branching ratio for the 
dimuon final state:
\be\label{SM:pred:BRbmumu}
\branch (\B_s\to \m^+\m^-)= (3.1 \pm 1.4) \times 10^{-9},
\ee 
where we have used the value $|V_{ts}|=0.04 \pm 0.002$ along with the 
aforementioned ranges for $f_{B_s},m_t$.
We emphasize that the error given is dominated by the uncertainty 
on the $B$ meson decay constant $f_{B_s}$.
Before moving on to the computation of Higgs-boson exchange diagrams that 
contribute to the form factors $F_i$, we briefly recall the experimental 
constraints relevant to our analysis. 

\subsection{Experimental constraints}\label{exp:constraints}
To date, the most stringent bounds on the magnitude of the  previously 
discussed short-distance coefficients  come from the  
Collider Detector at Fermilab (CDF) \cite{limits:exc:exp}:
\be\label{cdf:limit:btkstar}
\branch(B^0\to K^{*0} \m^+\m^-) < 4.0\times 10^{-6}\quad (\cl{90}),
\ee
which should be compared with the branching fraction
of about $2 \times 10^{-6}$  predicted by the SM. 
Also, from the absence of 
any signal from the process $B^+\to K^+ \m^+\m^-$, the $\cl{90}$ limit 
\be\label{cdf:limit:btok}
\branch(B^+\to K^+ \m^+\m^-) < 5.2 \times 10^{-6}
\ee
has been derived \cite{limits:exc:exp}, 
which is an order of magnitude away from the 
SM prediction of about $ 6 \times 10^{-7}$.

The measurement of the inclusive branching ratio $\branch(\B\to X_s \g)$ 
yields the result \cite{exp:btosg}
\be\label{range:btosg}
2.0\times 10^{-4}< \branch(\B\to X_s \g)<4.5\times 10^{-4}
\quad (\cl{95}),
\ee
which places limits on the absolute value of $\cseff$. In what follows it 
is more convenient to define the ratio 
$R_7\equiv \cseff/c_7^{\text{eff, SM}}$. 
Using the 
leading-order expression for $\branch(\B\to X_s \g)$ from  
\rf{kagan:neubert:btosg}, we calculate the bound to be 
\be\label{bsg:r7}
0.88 < |R_7| < 1.32.
\ee

A search for the decay $\B_s\to \m^+\m^-$ has been made by CDF, 
leading to the result \cite{exp:bstomumu}
\be\label{exp:limit:bsmumu}
\branch(\B_s\to \m^+\m^-)<2.6\times 10^{-6} \quad (\cl{95}).
\ee
This in turn translates, via \eq{BR:bll}, 
into an upper limit on the 
strength of scalar and pseudoscalar interactions, as we shall discuss. 

We conclude this section with a few remarks on the $\B$ mode. 
Experimental search leads to a 
$\cl{95}$ ($\cl{90}$) upper limit of 
$\branch(\B \to \m ^+\m^-)< 8.6  \times 10^{-7}$ \cite{exp:bstomumu}
($6.1  \times 10^{-7}$ \cite{exp:bdtomumu}), 
which is several orders of magnitude above the SM expectation 
of $O(10^{-10})$ \cite{Bsll:rec:rev}.~We stress that if flavour violation is 
due solely to the CKM matrix, the subject of the present paper, 
the $\B$ decay is suppressed relative to the $\B_s$ 
decay by a factor $|V_{td}/V_{ts}|^2\sim O(10^{-2})$; 
however, this 
suppression does not pertain to models with a new flavour structure.

\section{Higgs-boson contributions to \bm\lowercase{$b\to s l^+l^-$}}
\label{computation}
We now turn our attention to the computation of Higgs-boson contributions to 
the Wilson coefficients of the scalar and pseudoscalar 
operators in the $b\to s l^+l^-$ transition, within the context of the 2HDM and the
minimal supersymmetric standard model.\footnote{The Higgs-boson  
contributions to $\cseff, \ceff, \cten$ in the massless lepton 
approximation can be found in \rfs{bertolini:etal,charged:Higgs,frank:jorge}. 
For a non-zero lepton mass, 
there are also box diagrams with charged Higgs bosons which, at large 
$\tan\b$, contribute only to the helicity-flipped operators
${\Oi}_9'\sim (\bar{s} \gamma^{\mu} P_R b)(\bar{l}\gamma_{\mu} l)$ 
and ${\Oi}_{10}'\sim (\bar{s}\gamma^{\mu} P_R b)(\bar{l} \gamma_{\mu}\gamma_5 l)$ [cf.~\eq{operatorbasis}]; however, their contribution is  
negligible for $l=e$ or $\m$.}

As anticipated at the outset of this paper, we evaluate the diagrams in the 
$R_{\xi}$ gauge, which provides a check on the gauge invariance of our 
calculation. We use the  Feynman rules of \rf{fey:rosiek}
and focus on the large $\tan\b$ scenario, that is, 
$40\leqslant \tan\b\leqslant 60$.

\subsection{Two-Higgs-doublet model}
We compute the Higgs-boson exchange diagrams in the framework of a 2HDM 
where the up-type quarks couple to one Higgs doublet while the 
down-type quarks couple to the other Higgs doublet (usually referred to as 
model II), which occurs, for instance, in supersymmetry. We will use the SUSY constraints  
on the parameters $\la_i$ appearing in the Higgs potential (see, e.g., 
\rf{higgs:hunters}). We defer the discussion of the more 
general 2HDM with $\la_1=\la_2$, as well as the comparison with results
presented in the literature, to the end of the section. 

\begin{figure}
\begin{center}
\epsfig{file=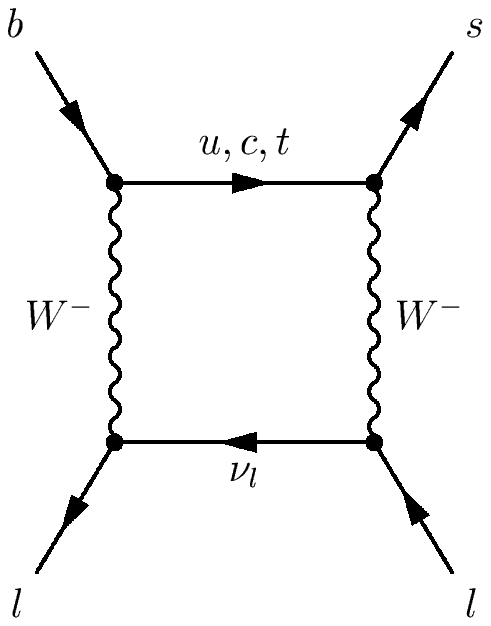,height=1.8in}\hspace{3em}
\epsfig{file=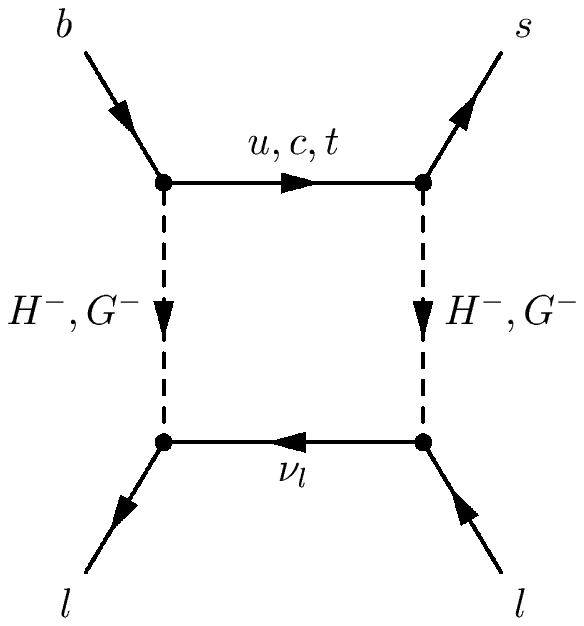,height=1.8in}\hspace{3em}
\epsfig{file=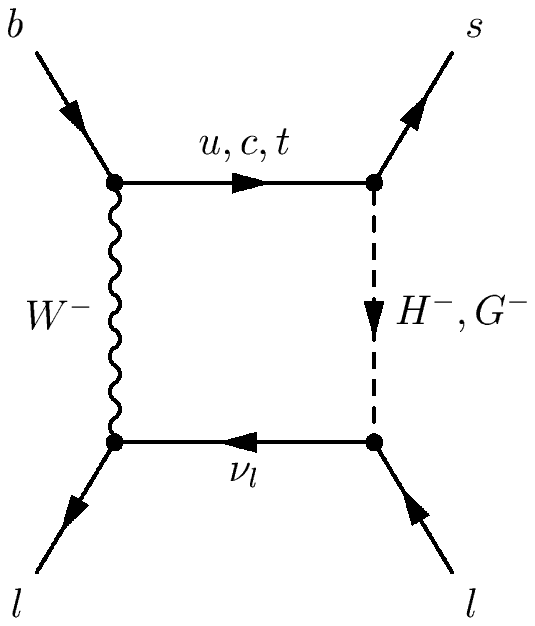,height=1.8in}\vspace{1.6em}
\epsfig{file=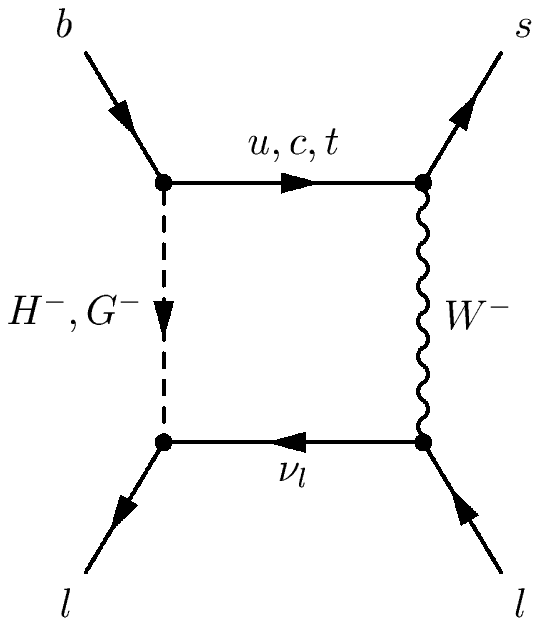,height=1.8in}\hspace{3em}
\epsfig{file=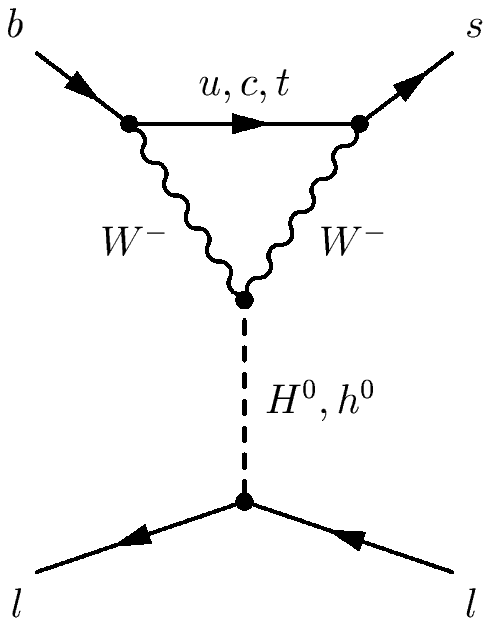,height=1.8in}\hspace{3em}
\epsfig{file=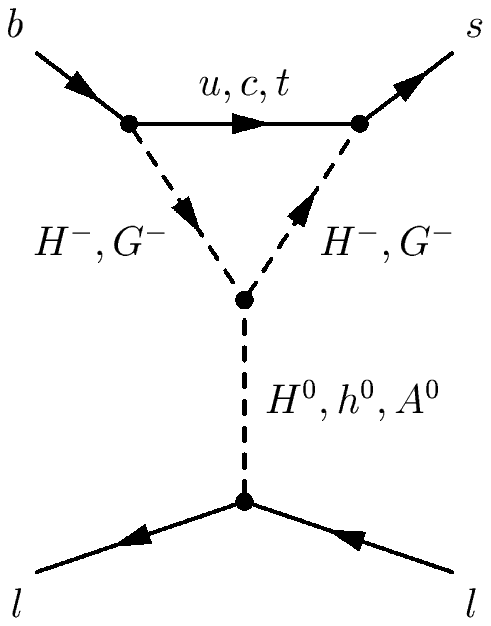,height=1.8in}\vspace{1.6em}
\epsfig{file=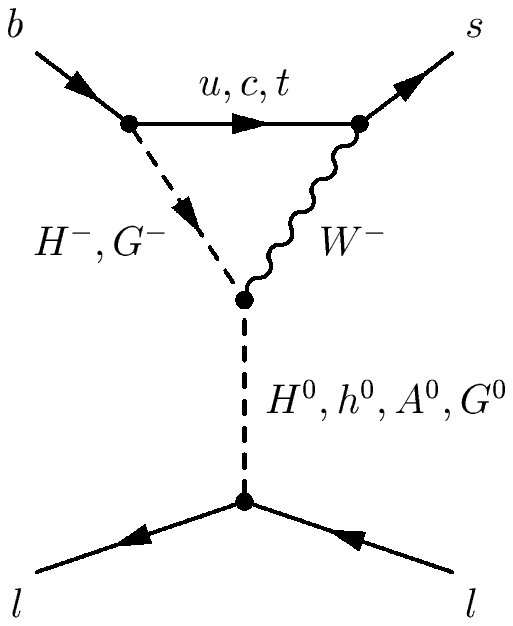,height=1.8in}\hspace{3em}
\epsfig{file=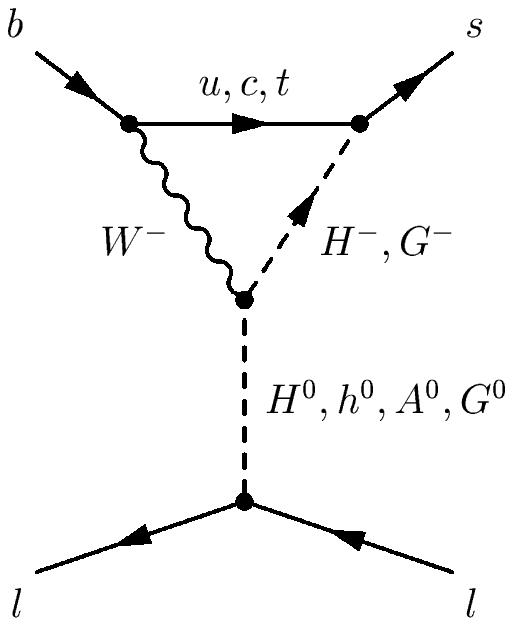,height=1.8in}\hspace{3em}
\epsfig{file=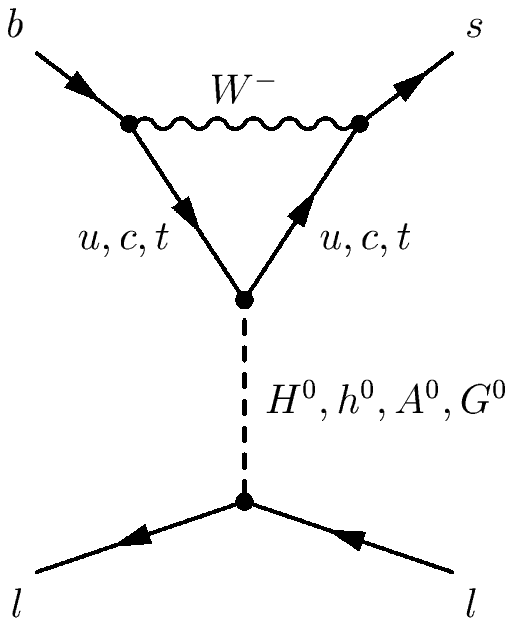,height=1.8in}\vspace{1.6em}
\epsfig{file=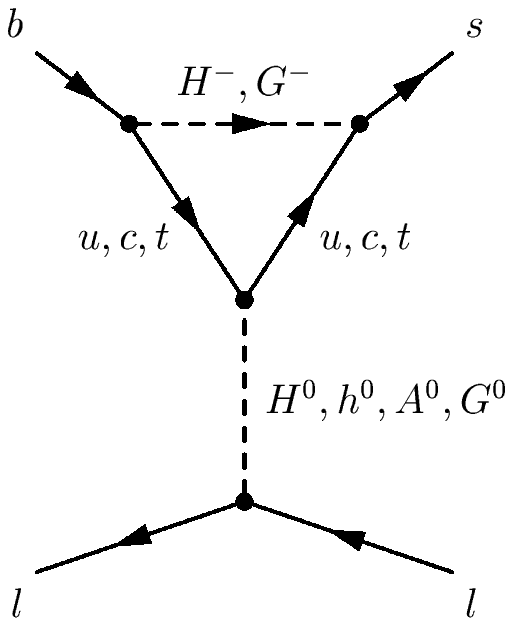,height=1.8in}\vspace{0.8em}
\caption{Diagrams contributing to the $b\to s l^+l^-$ transition, 
within the 2HDM. The corresponding counterterms are displayed in  
Eqs.~(\ref{counterterm:1}) and (\ref{counterterm:2}).\label{feyn:rules:2hdm}}
\end{center}
\end{figure}

The relevant Feynman diagrams for $b\to s l^+l^-$ 
are depicted in \fig{feyn:rules:2hdm}, where $A^0$  and $h^0,H^0$
are the $\cp$-odd and $\cp$-even Higgs bosons respectively, 
$H^\pm$ represents the charged Higgs bosons, and $G^0, G^\pm$ are the 
would-be-Goldstone bosons. 
%
% FIGURE 1
%
Before stating the results for the scalar and pseudoscalar Wilson 
coefficients, we pause to outline our renormalization procedure.

\subsubsection{Remarks on the renormalization procedure and the 
 renormalization group evolution}
Writing the interactions of Higgs bosons with down-type quarks 
appearing in the `bare' Lagrangian of the 2HDM and the minimal supersymmetric 
standard model (MSSM) in terms of renormalized quantities, we obtain 
the counterterms necessary to renormalize the theory,  
namely, in the one-loop approximation,  
\bea\label{counterterm:1}
\mbox{\raisebox{-1.5cm}[1.5cm]{\epsfxsize=3.5cm\epsffile{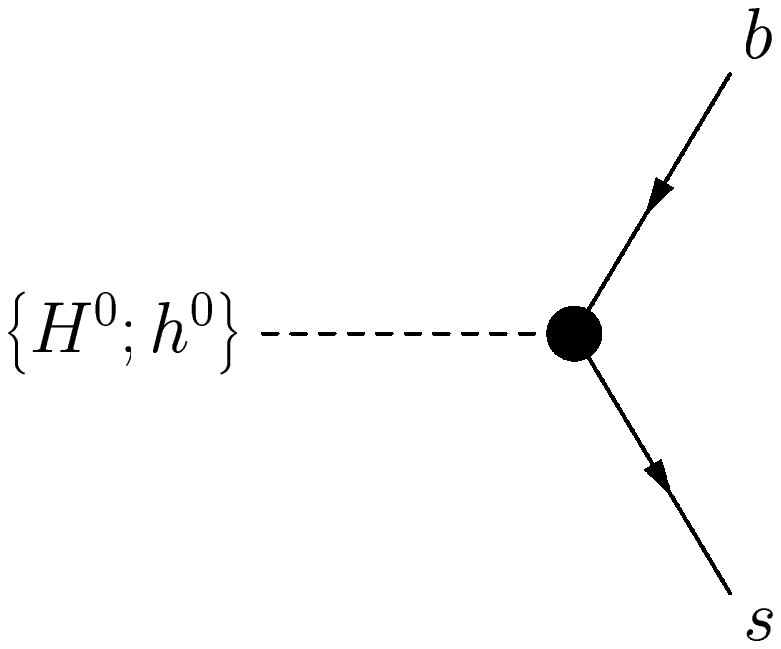}}}
\baR{c}\displaystyle
   -\frac{ie\left\{\cos\alpha;-\sin\alpha\right\}}{2M_W\sin\theta_W\cos\beta}
   \Bigg[\Bigg(\frac{m_s}{2}\delta Z_{sb}^R
   +\frac{m_b}{2}\delta Z^{L\dag}_{sb}\Bigg)P_R  \\
   \rule[0mm]{0mm}{8mm}\hspace{5cm}\displaystyle
   +\Bigg(\frac{m_s}{2}\delta Z_{sb}^L
   +\frac{m_b}{2}\delta Z^{R\dag}_{sb}\Bigg)P_L\Bigg],
\eaR
\eea
\bea\label{counterterm:2}
\mbox{\raisebox{-1.5cm}[1.5cm]{\epsfxsize=3.5cm\epsffile{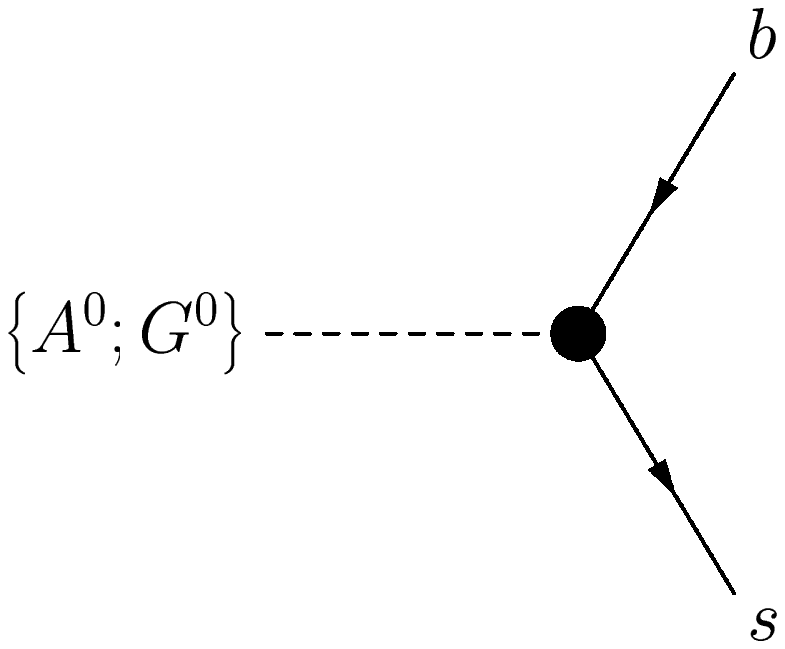}}}
\baR{c}\displaystyle
   -\frac{e\left\{\sin\beta;-\cos\beta\right\}}{2M_W\sin\theta_W\cos\beta}
   \Bigg[\Bigg( \frac{m_s}{2} \delta Z_{sb}^R
   +\frac{m_b}{2} \delta Z^{L\dag}_{sb} \Bigg)P_R \\
   \rule[0mm]{0mm}{8mm}\hspace{5cm}\displaystyle
   -\Bigg(\frac{m_s}{2}\delta Z_{sb}^L
   +\frac{m_b}{2}\delta Z^{R\dag}_{sb}\Bigg)P_L\Bigg],
\eaR
\eea
$\a$ being the mixing angle in the $\cp$-even Higgs sector,
and $\theta_W$ is the Weinberg angle.
The quark field renormalization constants are defined through
\be
 \left(\baR{c} d_L\\ s_L\\ b_L \eaR \right)_{\mathrm bare}
 =\left(\openone+\frac{1}{2}\delta Z^L\right)
 \left(\baR{c} d_L\\ s_L\\ b_L \eaR \right), \quad
 \left(\baR{c} d_R\\ s_R\\ b_R \eaR \right)_{\mathrm bare}
 =\left(\openone+\frac{1}{2}\delta Z^R\right)
 \left(\baR{c} d_R\\ s_R\\ b_R \eaR \right),
\ee
where $\openone$ is the unit matrix. They can be determined from the 
two-point function of the $\bar{s}b$ vertex, which is given by
\bea\label{two:point:function}
\mbox{\raisebox{-1.5cm}[1.5cm]{\epsfxsize=3.5cm\epsffile{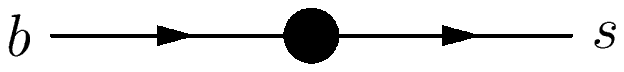}}}
\baR{c}\displaystyle
   \frac{i}{2}\bigg\{\bigg[\bigg(\delta Z_{sb}^R+\delta Z^{R\dag}_{sb} 
   \bigg)\pslash-\bigg(m_s\delta Z_{sb}^R
   +m_b\delta Z^{L\dag}_{sb}\bigg)\bigg]P_R \\
   \rule[0mm]{0mm}{8mm}\hspace{2cm}\displaystyle
   +\bigg[\bigg( \delta Z_{sb}^L+\delta Z^{L\dag}_{sb}\bigg)\pslash 
   - \bigg(m_s\delta Z_{sb}^L+m_b\delta Z^{R\dag}_{sb}\bigg)\bigg]P_L 
   \bigg\}.
\eaR
\eea
We have chosen an on-shell renormalization prescription in which the 
finite parts of the field renormalization constants are fixed by the 
requirement that the flavour-changing $b\to s$ vertex vanish for 
external on-shell fields.\footnote{Since our analysis is performed at
leading order in the electroweak couplings, the only quantity that 
has to be renormalized is 
the wave function.}~We have checked that our approach, based solely
upon one-particle-irreducible diagrams, yields a gauge-independent result, 
which is in agreement with \rf{logan:nierste}. 

Finally, a few remarks are in order regarding the renormalization group 
evolution. Referring to \eq{operatorbasis}, 
the masses of the light quarks, $m_{s}$ and $m_{b}$,
appear in the scalar and pseudoscalar operators
rather than in the corresponding Wilson coefficients, and hence  
must be evaluated at the low-energy scale. 
The anomalous dimensions of the light quark masses and the scalar as well 
as  pseudoscalar quark currents cancel, and so the anomalous dimension of 
the above-mentioned operators vanishes
(see also the discussion in \rf{chan:slaw}).
 Another method commonly used in the literature 
(see, e.g., \rfs{huang:etal:recent,chan:slaw}) is to absorb the light quark 
masses into the Wilson coefficients, which are determined at the 
high scale. In this case, the scalar and pseudoscalar operators have a 
non-vanishing 
anomalous dimension, and the values of the corresponding Wilson coefficients 
at the low scale are obtained by means of 
the renormalization group evolution. It is important to stress
that both methods are equivalent.

\subsubsection{Results for scalar and pseudoscalar couplings}
Retaining only leading terms in $\tan\b$, our results 
can be summarized as follows:
\be\label{2hdm:results:box}
c_S^{\Box} = -c_P^{\Box}=-\frac{m_l\tan^2\beta}{4M_W^2\sin^2\theta_W}
B(x_{H^{\pm}},x_t),
\ee
\be
c_S^{\Peng} = c_P^{\Peng} = 0,
\ee
\be\label{2hdm:results:count}
c_S^{\Count} = -c_P^{\Count}=-\frac{m_l\tan^2\beta}{4M_W^2\sin^2\theta_W}
C(x_{H^{\pm}},x_t),
\ee
where the superscripts denote the box-diagram, penguin, and counterterm 
contributions respectively, and 
$x_i = m_i^2/M_W^2$. The functions $B$ and $C$ are listed in 
Appendix \ref{aux:funcs}. In deriving 
\eqs{2hdm:results:box}{2hdm:results:count}, we have used the relation
\be\label{relation:tree-level}
\sin 2\a = -\sin2\b \left(\frac{M^2_{H^0}+M^2_{h^0}}{M^2_{H^0}-M^2_{h^0}}\right),
\ee
where $M_{H^0}, M_{h^0}$ 
are the  tree-level masses of the $\cp$-even Higgs bosons. 
Note that we have chosen $\tan\b$ and the charged Higgs-boson mass
$m_{H^{\pm}}$  as the two free parameters in this SUSY-inspired scenario.
Turning to the coefficients of the helicity-flipped operators $c_{S,P}'$,  
they are also proportional to $\tan^2\b$ but their contribution   
to the decay amplitude is suppressed by a factor of $m_s/m_b$ 
compared to $c_{S,P}$, and hence can be neglected.

Summing all contributions results in 
\be\label{constrained:2HDM:res}
c_S = -c_P = \frac{m_l\tan^2\beta}{4M_W^2\sin^2\theta_W}
\frac{\ln r}{1-r},\quad r=\frac{m_{H^{\pm}}^2}{m_t^2}.
\ee
(We will compare our findings  with other recent calculations below.)

\subsection{SUSY with minimal flavour violation}
\begin{figure}
\begin{center}
\epsfig{file=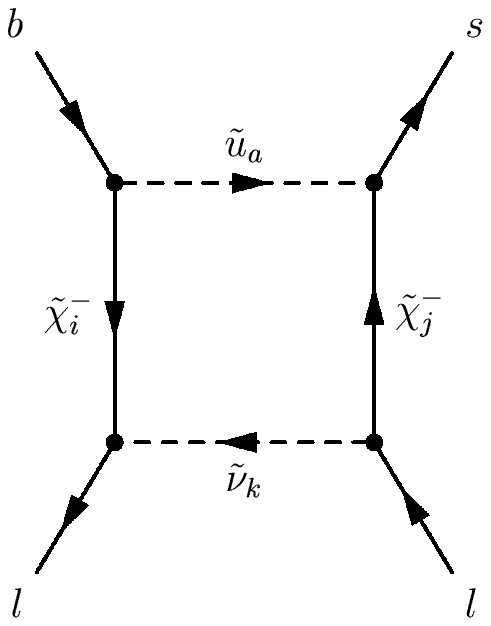,height=1.9in}\hspace{3em}
\epsfig{file=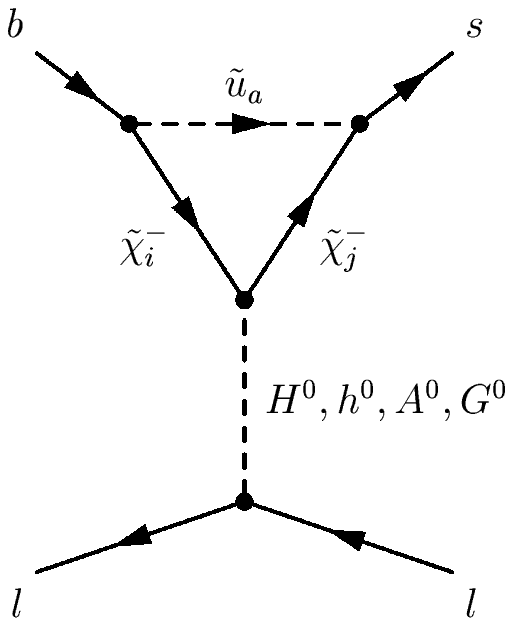,height=1.9in}\hspace{3em}
\epsfig{file=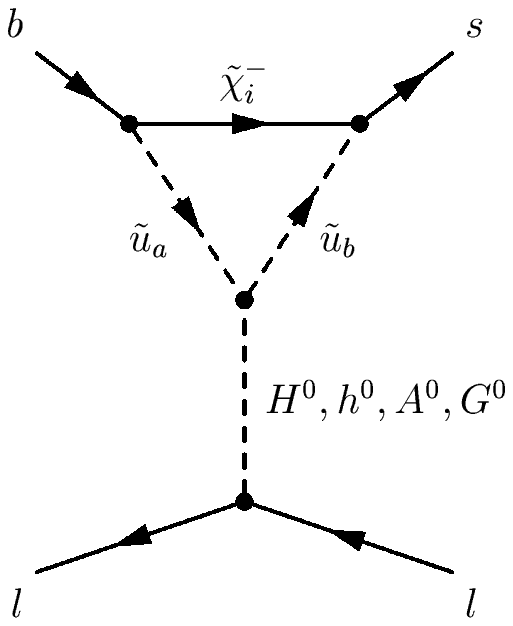,height=1.9in}\vspace{0.8em}
\caption{SUSY contributions to the process $b\to s l^+l^-$, 
as described in \sec{computation}, with indices 
$a=1,2,\dots,6$; $i,j=1,2$; $k=1,2,3$. The counterterms for the penguin diagrams are given in Eqs.~(\ref{counterterm:1}) and (\ref{counterterm:2}).\label{feyn:rules:susy}}
\end{center}
\end{figure}

Since we consider a scenario with minimal flavour violation, i.e.~we assume 
flavour-diagonal sfermion mass matrices, the contributing 
SUSY diagrams, in addition to those in \fig{feyn:rules:2hdm}, 
consist only of the two chargino states (see \fig{feyn:rules:susy}).\footnote{The gluino and neutralino 
contributions are deferred to another publication \cite{CTFJ:soon}.}
%
% FIGURE 2
%
It is convenient for the subsequent discussion to define the mass ratios
\be
   x_{ki}=\frac{m_{\sneutrino_k}^2}{m_{\tilde{\chi}_i^{\pm}}^2},\quad
   y_{ai}=\frac{m_{\upsquark_a}^2}{m_{\tilde{\chi}_i^{\pm}}^2},\quad
   z_{ij}=\frac{m_{\tilde{\chi}_i^{\pm}}^2}{m_{\tilde{\chi}_j^{\pm}}^2},
\ee
with $\sneutrino_k$, $\upsquark_a$, and $\chargino_i$ denoting sneutrinos, 
up-type squarks, and charginos. 
In terms of these variables and recalling \eq{relation:tree-level}, 
we obtain 
\bea\label{susy:result:box}
c_{S,P}^{\Box} = &\mp& \frac{m_l\tan^2\beta}
{2M_W}\sum_{i,j=1}^{2}\sum_{a=1}^{6}\sum_{k,m,n=1}^{3}
\frac{1}{m_{\tilde{\chi}_i^{\pm}}^2}
\Bigg\{
(R^\dagger_{\sneutrino})_{lk}(R_{\sneutrino})_{kl}
     (\Gamma^{U_L})_{am}U_{j2}\G^a_{imn}\nnu\\
   &\times&\Bigg(y_{ai} U_{j2}^{\ast}
V_{i1}^{\ast}\pm\frac{m_{\tilde{\chi}_j^{\pm}}}{m_{\tilde{\chi}_i^{\pm}}}
     U_{i2}V_{j1}\Bigg)D_1(x_{ki},y_{ai},z_{ji})\Bigg\},
\eea
\bea\label{susy:result:peng}
   c_{S,P}^{\Peng}  = &\pm&\frac{m_l\tan^2\beta}{M_W^2(m_{H^{\pm}}^2-M_W^2)}
\sum_{i,j=1}^{2}\sum_{a,b=1}^{6}\sum_{k,m,n=1}^{3}
     \G^a_{imn}(\Gamma^{U_L})_{bm}U_{j2}\nnu \\
   &\times&\Bigg\{M_W\Bigg(y_{aj}
     U_{j2}^{\ast}V_{i1}^{\ast}\pm\frac{m_{
     \tilde{\chi}_i^{\pm}}}{m_{\tilde{\chi}_j^{\pm}}}U_{i2}V_{j1}\Bigg)D_2(y_{aj},z_{ij})\delta_{ab}\delta_{km}\nnu\\
&-&\frac{ (M_U)_{kk}}{\sqrt{2}m_{\tilde{\chi}_i^{\pm}}}[\m^*(\Gamma^{U_R})_{ak}(\Gamma^{U_L\dagger})_{kb}
     \pm\m(\Gamma^{U_L})_{ak}(\Gamma^{U_R\dagger})_{kb}]
     D_2(y_{ai},y_{bi})\delta_{ij}\Bigg\},
\eea
\bea\label{susy:result:count}
c_{S,P}^{\Count} &=&\mp
\frac{m_l\tan^3\beta}{\sqrt{2}M_W^2
(m_{H^{\pm}}^2-M_W^2)}\sum_{i=1}^{2}\sum_{a=1}^{6}\sum_{m,n=1}^{3}
 [m_{\tilde{\chi}_i^{\pm}} D_3(y_{ai})U_{i2}(\Gamma^{U_L})_{am}\G^a_{imn}],
\eea
where  
\be
M_U\equiv \diag(m_u, m_c, m_t),
\ee
\be\label{susy:result:gamma}
\G^a_{imn}= \frac{1}{2\sqrt{2}\sin^2\theta_W}
[\sqrt{2}M_W V_{i1}(\Gamma^{U_L\dagger})_{na}-(M_U)_{nn}V_{i2}
     (\Gamma^{U_R\dagger})_{na}]\la_{mn},
\ee
with the ratio of CKM factors 
$\la_{mn}\equiv V_{mb}^{}V_{ns}^{\ast}/V_{tb}^{}V_{ts}^{\ast}$, and the 
functions $D_{1,2,3}$ are listed in Appendix 
\ref{aux:funcs}. In writing the above formulae, we have used the unitarity 
of the CKM matrix and the squark mixing matrices 
(for definitions see Appendix \ref{app:susy:mass-matrices}).
Note that the chargino contributions vanish when all the 
scalar masses are degenerate, reflecting the unitarity of the mixing 
matrices. Another noticeable feature is that the leading term in $\tan\b$ 
comes from the counterterm diagrams.

Turning to the Wilson coefficients of the helicity-flipped scalar and 
pseudoscalar operators entering the operator basis 
in \eq{operatorbasis}, we obtain 
\be\label{cis:prime}
c_S^{\prime\,\Count}=[c_S^{\Count}(\la_{mn}\to \la_{nm}^\ast)]^\ast, \quad 
c_P^{\prime\, \Count}=-[c_P^{\Count}(\la_{mn}\to \la_{nm}^\ast)]^\ast,
\ee
which are $O(\tan^3\b)$, while the remaining coefficients are proportional 
to $\tan^2\b$. Recall that the 
contribution of the Wilson coefficients in \eq{cis:prime} to the decay 
amplitude is proportional to $m_s\tan^3\b$ and hence comparable in size to 
the $m_b\tan^2\b$ contributions, Eqs.~(\ref{constrained:2HDM:res}), 
(\ref{susy:result:box}), (\ref{susy:result:peng}). 
On the other hand, this contribution
is negligible when compared with the leading term $m_b\tan^3\b$ and 
so will be omitted in what follows.

\subsection{Remarks on previous results}
We conclude this section by comparing our results with 
previous calculations in the 
literature \cite{logan:nierste,huang:etal:recent,chan:slaw}. 
To this end, we consider the MSSM as well as a general 2HDM with 
$\la_1=\la_2$ for the coupling constants 
appearing in the Higgs potential \cite{higgs:hunters}. 
We discuss  the two scenarios in turn.

(a) Working in the 
framework of the MSSM, the Higgs sector is equivalent to the one 
of the 2HDM with SUSY constraints \cite{higgs:hunters}. 
The results of Huang 
\ea\ \cite{huang:etal:recent} 
can be checked by exploiting  the tree-level relation
in \eq{relation:tree-level}. Reducing their expressions for the 
one-loop functions to the compact formulae given above, and after correcting 
numerous typographical errors, we agree with their results. 
Note that our anomalous 
dimension is equal to zero, due to the running $b$-quark mass 
entering the definition of the operators [see \eq{operatorbasis}].
In order to compare our findings with those obtained by Chankowski 
and S\l awianowska \cite{chan:slaw}, we specialize to the case 
$M_2 \gg |\m|$, so that
$m_{\chargino_1}\approx |\m|$ and $m_{\chargino_2}\approx |M_2|$.
In this approximation, and retaining only contributions of the lighter 
chargino and the scalar top quark, 
our results are in agreement with Eqs.~(33) and (34) of \rf{chan:slaw}.

(b) In the context of the general $\cp$-conserving 2HDM with the constraint 
$\la_1=\la_2$, the set of free 
parameters consists of $M_{h^0}, M_{H^0}, M_{A^0}, m_{H^{\pm}}$, as well as 
the 
mixing angles $\a$ and $\b$. 
The necessary Feynman rules are listed in \rf{higgs:hunters} apart 
from the model-dependent trilinear Higgs couplings $g_{H^+H^-h^0}$ and 
$g_{H^+H^-H^0}$, which can be found, for example, in 
\rf{skiba:kali}. (We have rederived these 
couplings confirming the result given there.) In the limit of large $\tan\b$, these 
couplings reduce to 
\be\label{tri:h0}
g_{H^+H^-h^0}\simeq -\frac{i g}{2M_W}\sin\a \cos^2\a (M_{H^0}^2-M_{h^0}^2)
\tan\b[1+ O(\cot\b)],
\ee  
\be\label{tri:H0}
g_{H^+H^-H^0}\simeq -\frac{i g}{2M_W}\sin^2\a \cos\a(M_{H^0}^2-M_{h^0}^2)
\tan\b[1+ O(\cot\b)].
\ee 
Our results for the trilinear couplings disagree with those 
presented in Eq.~(27) of \rf{logan:nierste}. In addition, we would 
like to stress 
that within the general 2HDM the mixing angles $\a$ and $\b$ are 
independent parameters, contrary to the statement made in 
that work. (This has also been pointed out in 
\rf{huang:etal:recent}.)
Taking into account the Feynman diagrams due to the trilinear
Higgs couplings in Eqs.~(\ref{tri:h0}) and (\ref{tri:H0}), 
we obtain 
\be\label{2hdm:general:cs}
c_S = \frac{m_l\tan^2\beta}{4M_W^2\sin^2\theta_W}\Bigg\{
\frac{\ln r}{1-r} + \sin^2(2 \a) \frac{(M_{H^0}^2-M_{h^0}^2)^2}{4M_{H^0}^2 M_{h^0}^2}\Bigg[\frac{1-r +\ln r}{(1-r)^2}\Bigg]\Bigg\}, 
\quad r=\frac{m_{H^{\pm}}^2}{m_t^2},
\ee
and $c_P$ as given in \eq{constrained:2HDM:res}. 
The first term is in agreement with the calculation of Logan and Nierste 
\cite{logan:nierste} and with the result obtained by Huang \ea\ 
\cite{huang:etal:recent}. As for the $\a$-dependent term, it is absent
in the expression given in \rf{logan:nierste} and differs from that of
\rf{huang:etal:recent}. We caution 
that in models such as  the general 
2HDM with a complicated parameter space the subleading terms in $\tan\b$ might be of the same order of magnitude 
as the leading ones for certain values of the parameters.   
%
%%%%%%%%%%%%%%%%%%%%%%%%%%%%%%%%%%%%%%%%%%%%%%%%%%%%%%%%%%%%%%%%%%%%%%%%%%%%%
%
%In addition, we would like to point out that  
%in general there is no reason to expect the $\tan^2\b$ term in 
%\eq{2hdm:general:cs} to be the dominant one. As a matter of fact, the above 
%Wilson coefficient has the structure  
%\be
%c_S= a\tan^2\b+ b\tan\b +c,
%\ee
%wher $a,b,c$ are non-trivial functions of the aforementioned parameters in 
%the general 2HDM. Thus, in order to assess the importance of the 
%$\tan^2\b$ term, a complete calculation to all orders in $\tan\b$ must 
%be performed.
%
%%%%%%%%%%%%%%%%%%%%%%%%%%%%%%%%%%%%%%%%%%%%%%%%%%%%%%%%%%%%%%%%%%%%%%%%%%%%%
%
\section{Implications for the  decays 
\bm $\B_{\lowercase{s}}\to \m^+\m^-$ and  
\bm $\B\to K \m^+\m^-$}\label{num:analysis}
In this section we explore the consequences of the current upper limits 
on rare $B$ decays discussed in \sec{dec:dis} for scalar and pseudoscalar 
interactions. In the quantitative analysis, we use 
$m_b\equiv \ol{m}_b(m_b) = 4.4\ \GeV$ and neglect 
terms of order $m_s/m_b$, which is certainly sufficient for our purposes. 

\subsection{Model-independent analysis}
We start by analysing the  constraints on scalar and pseudoscalar 
interactions, as well as on $B$-physics observables, in a model-independent 
manner. To this end,
we define the following dimensionless quantities
\be\label{def:Ris}
R_i\equiv c_i/c_i^{\sm}, \quad R_{S,P}\equiv m_b c_{S,P},
\ee
with $c_i^{\sm}$ as in \eq{coeffs:sm}. For the purposes of this analysis, 
we assume that there are no additional $\cp$ phases, 
besides the single CKM phase, 
so that the $R$'s in \eq{def:Ris} are real (remembering that we omit 
terms proportional to $V_{ub}^{}V^*_{us}/V_{tb}^{}V^*_{ts}\ll 1$).

\subsubsection{Bounds on scalar and pseudoscalar couplings}
\begin{figure}
\begin{center}
\epsfig{file=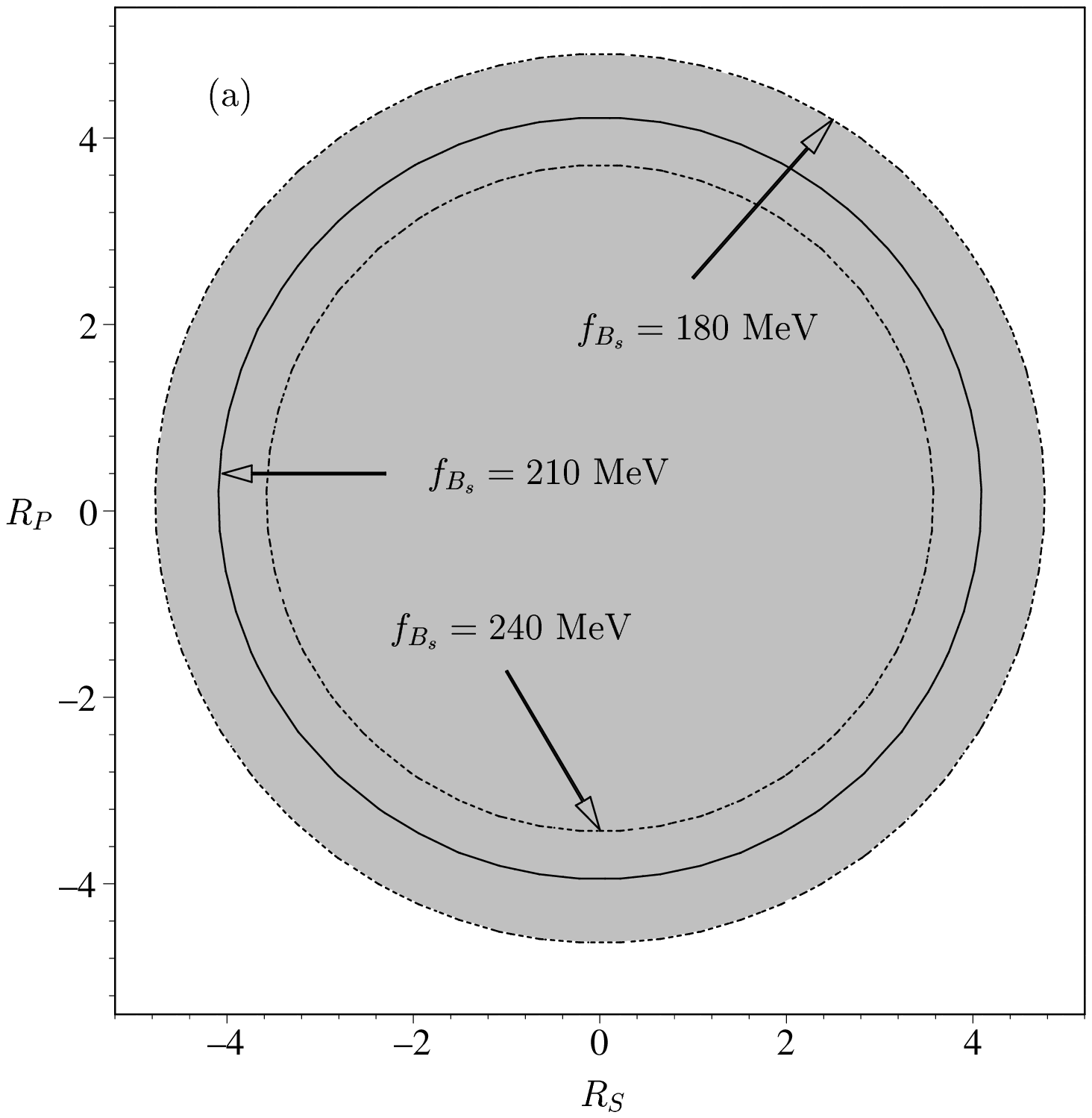,height=3.2in}\hspace{1em}
\epsfig{file=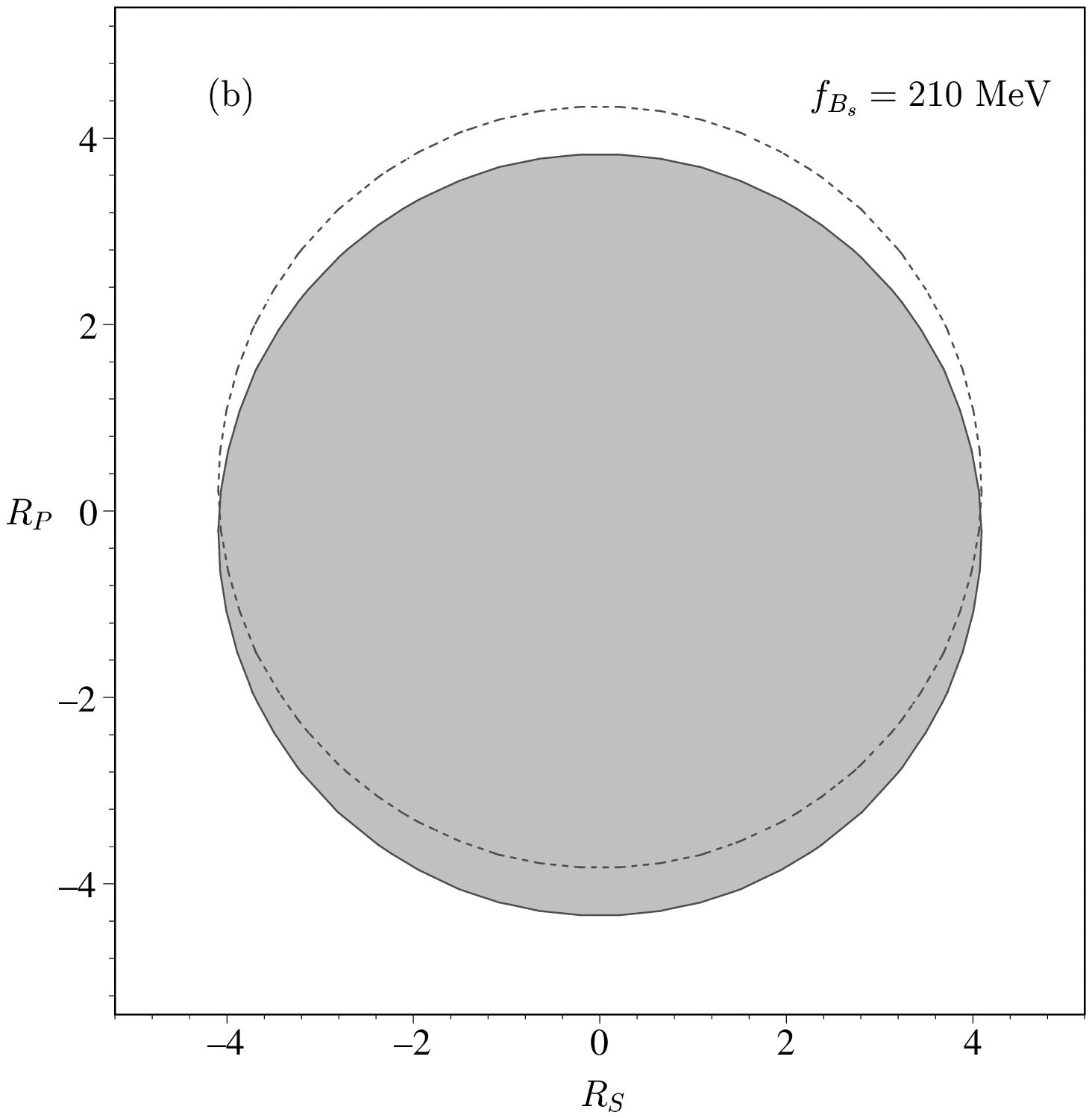,height=3.2in}\vspace{0.8em}
\caption{Constraints on the coefficients $R_S$ and $R_P$ as determined 
from the upper limit on $\branch(\B_s\to \m^+\m^-)$. 
(a) For $R_{10}=1$ and  $f_{B_s}=210 \pm 30\ \MeV$.
The dark region indicates the allowed region.
(b) For the central value $f_{B_s}=210\ \MeV$ and two choices of $R_{10}$: 
$-2$ (solid line) and $2$ (dashed line), as described in 
the text.\label{constraints:cscp}}
\end{center}
\end{figure}

The most severe constraints on scalar and pseudoscalar 
interactions arise, as we will argue shortly, 
from the upper bound on the $\B_s\to \m^+\m^-$ 
branching fraction, \eq{exp:limit:bsmumu}, which  maps out an allowed 
region in the $(R_S,R_P)$ plane. This is illustrated 
in \fig{constraints:cscp}(a), where we have chosen 
$f_{B_s}=210\pm 30\ \MeV$ and assumed 
$R_{10}=1$ (i.e., the SM value for $\cten$). 
We note that the allowed region in the $(R_S, R_P)$ plane is fairly 
insensitive to the range $-2 \lesssim R_{10} \lesssim 2$ implied 
by the present experimental bound on 
$\branch(\B\to K^* \m^+\m^-)$.
This can be easily understood from \eq{BR:bll},
where the contribution of $R_{10}$, or equivalently of $c_{10}$, 
to the branching ratio is helicity suppressed.
It is important to emphasize that the 
maximum allowed contribution of scalar and pseudoscalar operators to the 
$\B\to K^* \m^+\m^-$ branching fraction is consistent with the experimental 
upper limit, \eq{cdf:limit:btkstar}. As will become clear, 
the new-physics contribution due to 
Higgs-mediated interactions does not  significantly alter the maximum
allowed values of $R_9$ and  $R_{10}$.\footnote{The bounds on $R_9$ 
also depend on the sign and magnitude of $R_7$. For details, see 
Figs.~9 and 10 in \rf{ali:etal}, and \fig{constraints:r9r10} below.}  

Taking $R_{10}=\pm 2$ and $f_{B_s}=210\ \MeV$, 
we infer from \fig{constraints:cscp}(b) the interval
$-4 \lesssim R_{S,P}\lesssim 4$ for scalar and pseudoscalar couplings.
%
% FIGURE 3
%
\subsubsection{Branching ratio and FB asymmetry in $\B\to K \m^+\m^-$}
\begin{figure}[p]
\begin{center}
\epsfig{file=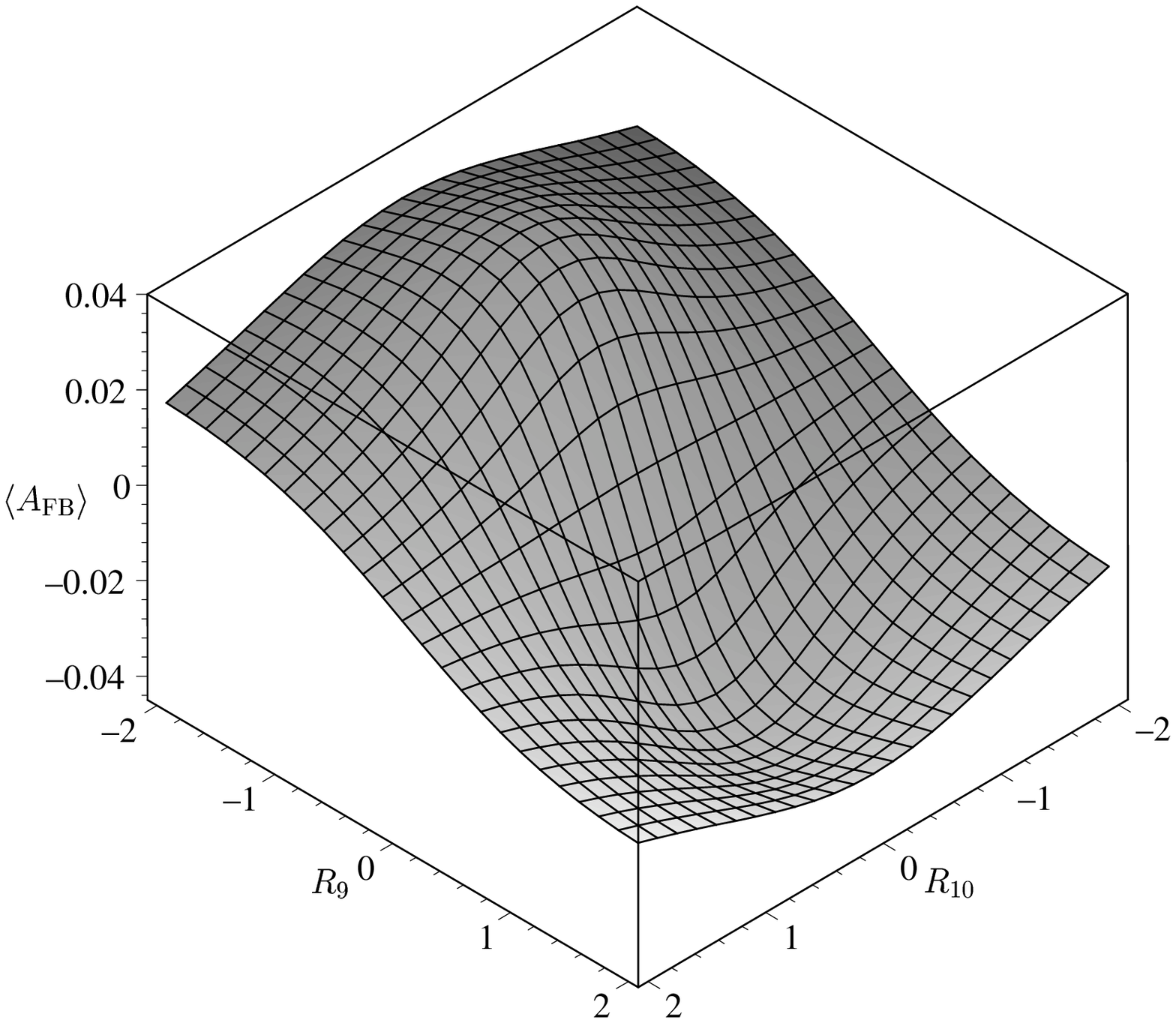,height=2.8in}
\epsfig{file=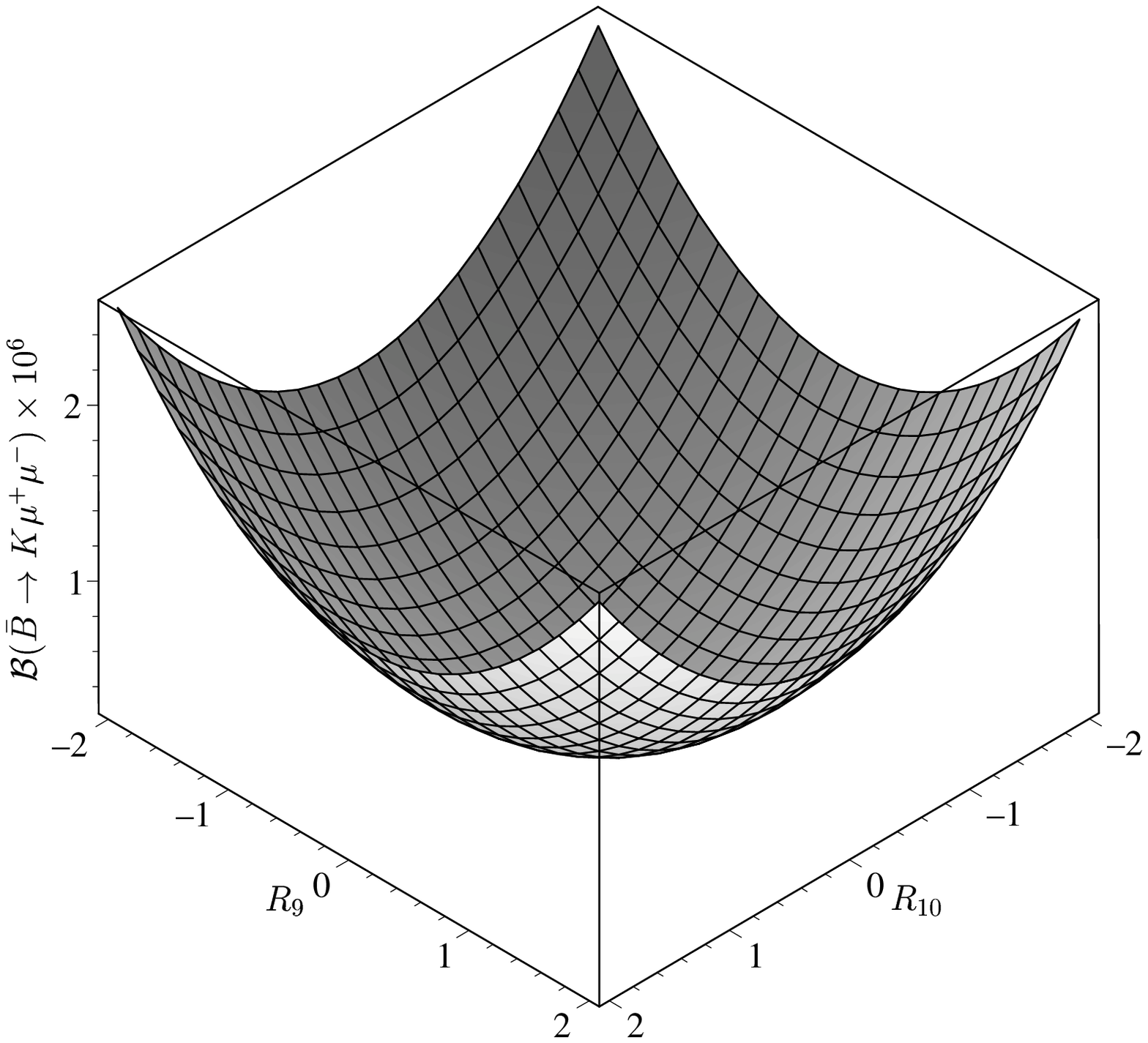,height=2.8in}\vspace{0.8em}
\caption{The average FB asymmetry in $\B\to K\m^+\m^-$ as a function of
$(R_9,R_{10})$ for  
$R_S=-4$, $R_P=0$, and $R_7=1$ consistent with the upper limit on 
$\branch(\B_s\to \m^+\m^-)$. Also shown is the corresponding 
branching ratio of $\B\to K \m^+\m^-$ (see the text for details).\label{fb:asymmetry:rsmax}}
\end{center}
\end{figure}

\begin{figure}
\begin{center}
\epsfig{file=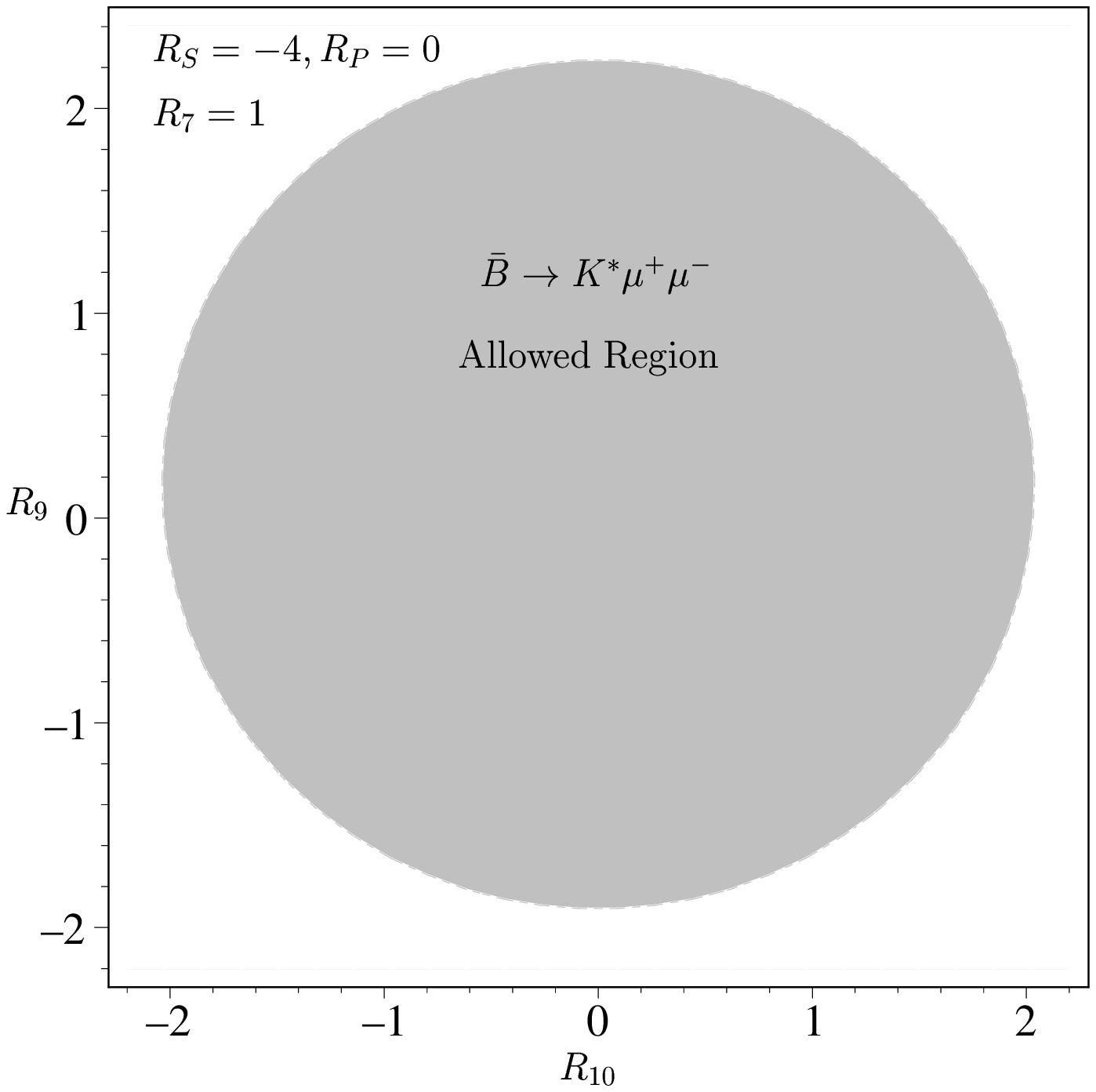,height=3in}\hspace{1em}
\epsfig{file=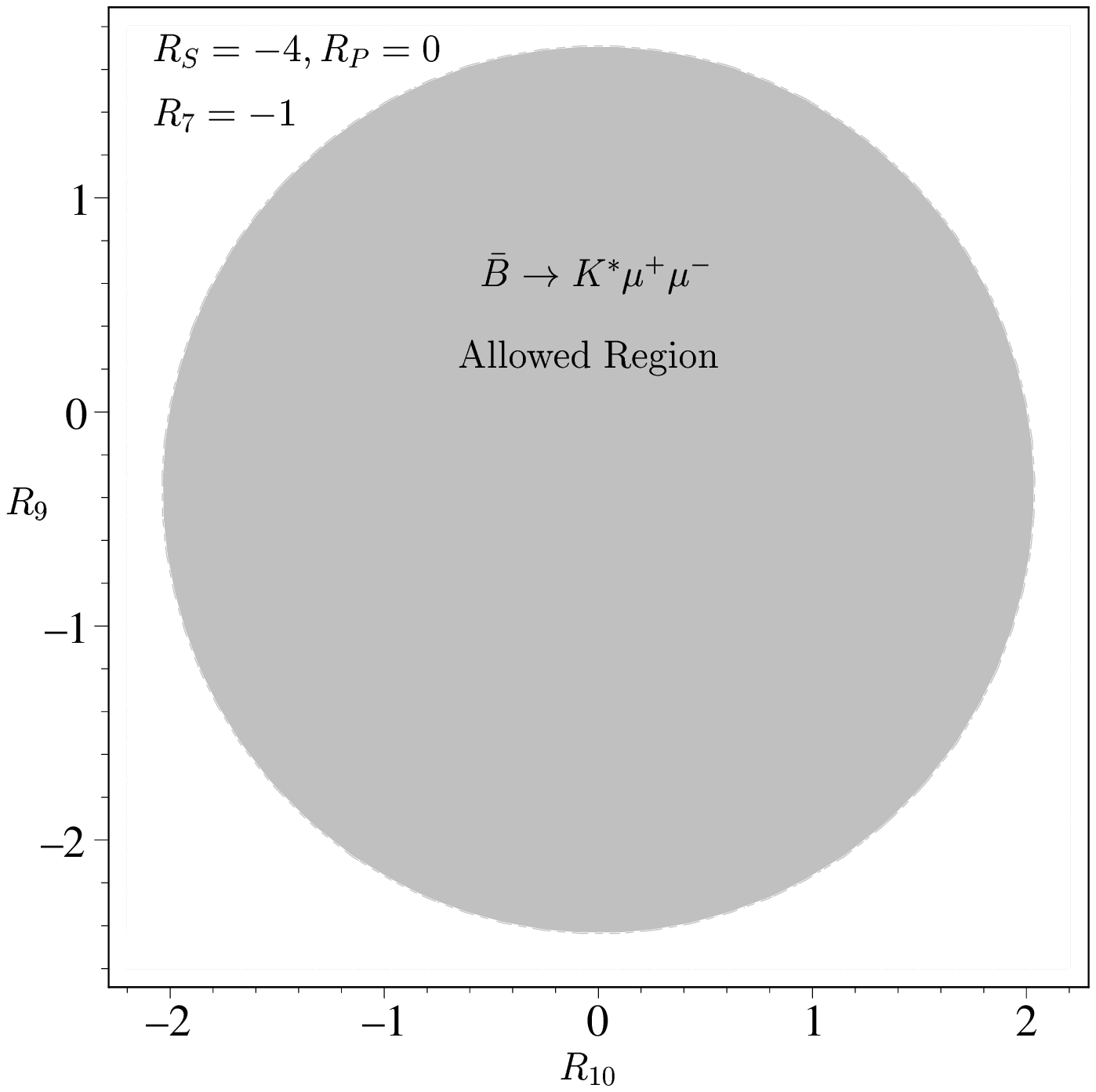,height=3in}\vspace{0.8em}
\caption{Allowed ranges of $R_9$ and $R_{10}$ as determined from 
the upper limit on $\branch(\B\to K^*\m^+\m^-)$ for $R_S=-4$, $R_P=0$, and 
$R_7=\pm 1$. 
Note that the values of $R_{S,P}$ are consistent with experimental 
data on $\B_s\to \m^+\m^-$.\label{constraints:r9r10}}
\end{center}
\end{figure}

We now assess the implications of scalar and pseudoscalar 
interactions for the branching ratio and forward-backward asymmetry.
For the paper to be self-contained, we also provide the analytic 
expression for the $\B\to K^* \m^+\m^-$ branching fraction.
If we keep in mind that the Wilson coefficients are real, we obtain 
\bea
\branch(\B \to K\m^+\m^-)&=& [0.042+ 
2.846 R_{10}^2 + 2.730 R_9^2 + 0.043 R_7^2 + 0.181 R_S^2 + 0.182 R_P^2 \nnu \\
&-&0.132 R_P R_{10} - 0.686 R_7 R_9 
+ 0.522 R_9  - 0.065 R_7]\times 10^{-7},
\eea
\bea\label{BR:BtoKstar:expression}
\branch(\B \to K^*\m^+\m^-)&=& [0.015+ 
0.922R_{10}^2 + 0.890 R_9^2 + 0.212 R_7^2 + 0.014 R_S^2 + 0.014 R_P^2 \nnu \\
&-&0.015 R_P R_{10} - 0.469 R_7 R_9 
+ 0.177 R_9  - 0.046 R_7]\times 10^{-6},
\eea
where the $R$'s are defined in \eq{def:Ris}. We note that the limits on 
$R_S$ and $R_P$ from the upper bound on 
$\branch(\B\to K \m^+\m^-)$ [\eq{cdf:limit:btok}] are numerically less 
stringent than those derived previously from $\B_s\to \m^+\m^-$.
Consequently, $|R_{S,P}| \simeq 4$ is essentially the largest value 
that is possible. Some representative results for the branching ratios of 
$\B_s\to\m^+\m^-$ and $\B\to K^{(*)}\m^+\m^-$ are summarized in Table 
\ref{table:num:res:rs:rp}. 
% 
%%%%%%%%%%%% Table 1 %%%%%%%%%%%%%%%%%
%
\begin{table}
\caption{Branching ratios of the decays $\B_s\to\m^+\m^-$
and $\B\to K^{(*)}\m^+\m^-$ for the choice $(R_7,R_9,R_{10})=(-1.2,1.1,0.8)$
together with the present experimental upper limits (see \sec{dec:dis}).\label{table:num:res:rs:rp}}
\begin{tabular}{lccc}
$(R_S, R_P)$ & $\branch(\B_s\to\m^+\m^-)$&
$\branch(\B\to K\m^+\m^-)$& 
$\branch(\B\to K^*\m^+\m^-)$\\
\hline \vspace{-1em}\\
$(0,0)$ &$1.8\times 10^{-9}$ &$6.8\times 10^{-7}$&$2.8\times 10^{-6}$  \\
$(2,2)$ &$1.1\times 10^{-6}$ &$8.0\times 10^{-7}$&$2.8\times 10^{-6}$ \\
$(3,-2)$ &$2.1\times 10^{-6}$ &$9.3\times 10^{-7}$&$2.8\times 10^{-6}$ \\
$(-4,0)$ &$2.5\times 10^{-6}$ &$9.7\times 10^{-7}$&$2.8\times 10^{-6}$ \\
\hline
Exptl  limits  &$<2.6\times 10^{-6}\ (\cl{95})$ &
$<5.2\times 10^{-6}\ (\cl{90})$&
$<4.0\times 10^{-6}\ (\cl{90})$\\
\end{tabular}
\end{table}
As can be seen, the branching ratio of $\B\to K^* \m^+\m^-$ decay is 
essentially unaffected by the presence of Higgs-mediated interactions
since the contributions of $R_{S,P}$ in \eq{BR:BtoKstar:expression} are 
suppressed compared to those of $R_{9,10}$. 
Furthermore, 
it is important to emphasize that large effects of scalar and 
pseudoscalar interactions on the 
$\B\to K \m^+\m^-$ decay rate are already excluded by the 
upper limit on the $\B_s\to \m^+\m^-$ branching fraction.~(This 
constraint has not been taken into account in the analysis of 
\rf{huang:etal:bkmumu}.)
As for the asymmetry, our main interest is in 
the average FB asymmetry, which can be obtained from the expression in 
\eq{FB:explicitly} by integrating numerator and denominator separately over 
the dilepton invariant mass, leading to  
\be
\av{A_{\text{FB}}}=R_S(0.512 + 5.424 R_9 -0.681 R_7) 
\Bigg(\frac{10^{-9}}{\branch}\Bigg).
\ee
To gain a maximum FB asymmetry, we fix
$R_S=-4$ and $R_P=0$ allowed by current experimental data on  
$\branch(\B_s\to \m^+\m^-)$, together with the SM value of $R_7=1$. 
Referring to \fig{fb:asymmetry:rsmax}, 
%
% FIGURE 4
%
it is evident that the average FB asymmetry in $\B\to K \m^+\m^-$ decay 
amounts to $\pm 4\%$ at most, the actual value depending on $R_9$ and $R_{10}$.
We emphasize that some of the values of $(R_9,R_{10})$, 
while respecting the upper bound on  $\branch(\B\to K \m^+\m^-)$,
are not compatible with the experimental constraint 
on the $\B\to K^* \m^+\m^-$ 
branching fraction. This is illustrated in 
\fig{constraints:r9r10}, where we show  the allowed range of $R_{9,10}$.
%
% FIGURE 5
%
Table \ref{table:num:res:gen} contains our predictions for the maximum 
average FB asymmetry and the branching ratios of $\B\to K^{(*)} \m^+\m^-$
for certain choices of parameters.
%
%%%%%%%%%%%% Table 2 %%%%%%%%%%%%%%%%%
%
\begin{table}
\caption{Maximum values of the 
average FB asymmetry in 
$\B\to K\m^+\m^-$ decay together with the branching ratios of 
$\B\to K^{(*)}\m^+\m^-$ for different choices of $(R_7,R_9,R_{10})$. We have 
chosen $(R_S,R_P)=(-4,0)$, resulting in 
$\branch(\B_s\to \m^+\m^-)\simeq 2.5\times 10^{-6}$, which is close to 
the present upper bound of $2.6\times 10^{-6}$. Also listed are 
the $\cl{90}$ upper limits as discussed in \sec{dec:dis}. 
\label{table:num:res:gen}}
\begin{tabular}{lccc}
$(R_7,R_9,R_{10})$ &$\av{A_{\text{FB}}}$& $\branch(\B\to K\m^+\m^-)$& 
$\branch(\B\to K^*\m^+\m^-)$\\
\hline \vspace{-1em}\\
$(1, 1.9,1)$ &$-2.6\%$ &$1.5\times 10^{-6}$ & 
$3.8\times 10^{-6}$\\
$(-1,1.2, 1.3)$ &$-2.3\%$&$1.3\times 10^{-6}$ & 
$3.9\times 10^{-6}$\\
$(1,1,1)$ &$-2.5\%$  &$8.3\times 10^{-7}$ & 
$1.7\times 10^{-6}$\\
$(1,-1,0)$ &$+3.8\%$ &$5.8\times 10^{-7}$ & 
$1.4\times 10^{-6}$\\
$(1,1,0)$ &$-3.8\%$ &$5.5\times 10^{-7}$ & 
$7.8\times 10^{-7}$\\  
$(-1.2,1.1,0.8)$ &$-3.0\%$ &$9.7\times 10^{-7}$ & 
$2.8\times 10^{-6}$\\ \hline 
Exptl limits & $-$ &$<5.2\times 10^{-6}$&
$<4.0\times 10^{-6}$\\
\end{tabular}
\end{table}
Note that the measurement of a nominal asymmetry of $4\%$ (at $3\s$ 
level), which is accompanied by a branching fraction of 
$\sim 6 \times10^{-7}$, 
will  necessitate at least $\sim 10^{10}$
$B$ mesons and could conceivably be measured
in forthcoming 
experiments at the CERN Large Hadron Collider (LHC) and the Tevatron.
We conclude that the predicted FB asymmetry due to scalar interactions, 
though larger than in the SM, may possibly be too small
to be seen experimentally. Nevertheless, the FB asymmetry does provide  
a very useful laboratory for studying possible extensions of the SM.

\subsection{Constraints on new physics with minimal flavour violation}
It is clear from the previous analysis that the upper bound on the 
$\B_s\to \m^+\m^-$ branching fraction gives the strongest constraints on the 
scalar and pseudoscalar couplings, $c_{S,P}$,
which in turn can be used to restrict the parameter space of models 
outside  the SM. 

%We also examine the new-physics contribution due to Higgs-boson 
%exchange to the Wilson coefficients $\cseff,\ceff,\cten$.

\subsubsection{Two-Higgs-doublet model}
\begin{figure}
\begin{center}
\epsfig{file=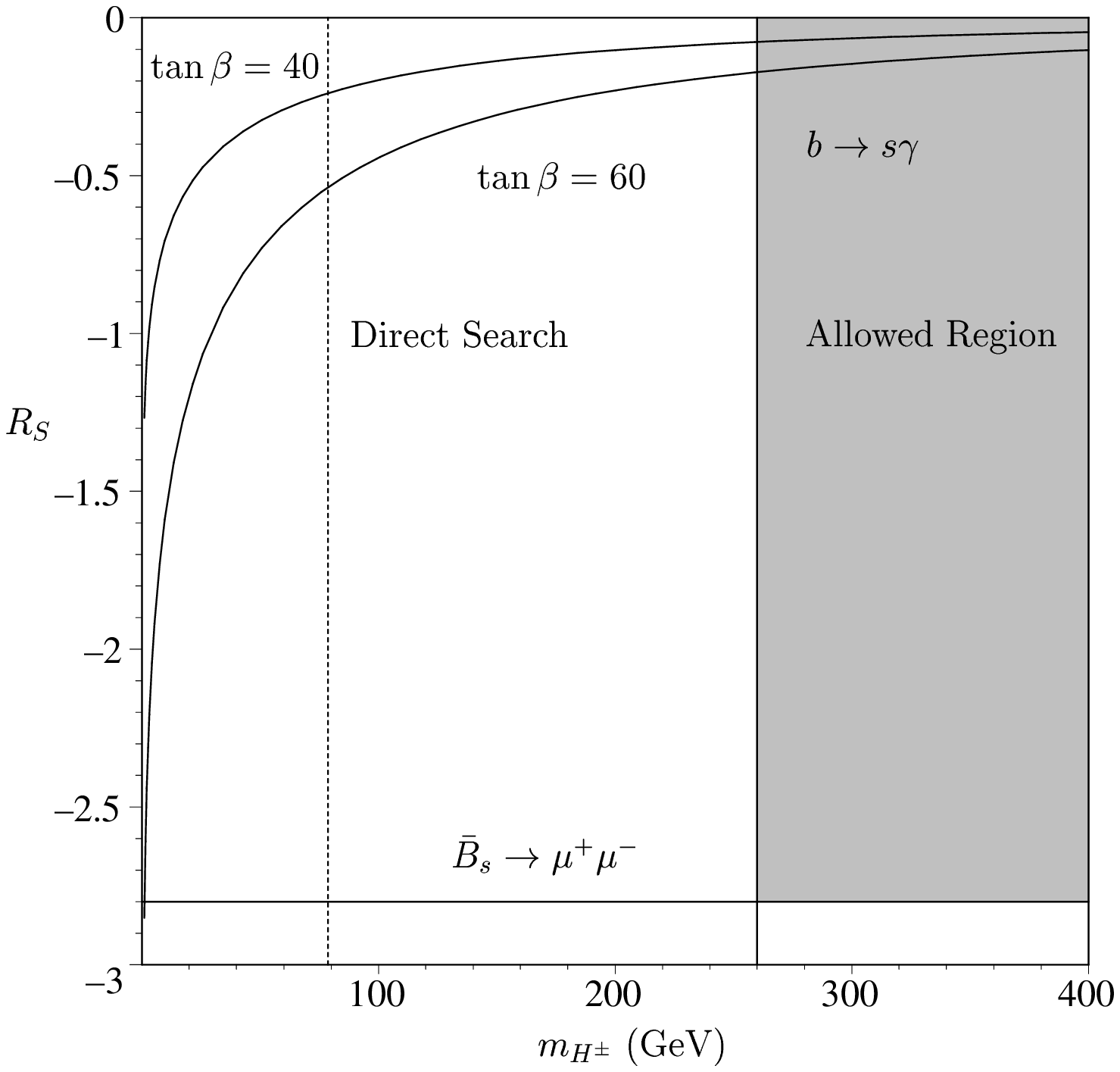,height=3.1in}\vspace{0.8em}
\caption{$R_S$ vs the charged Higgs-boson mass in the 
2HDM for large values of $\tan\b$, taking into  account the constraint from 
$\B_s\to \m^+\m^-$ decay. We exhibit the cases of $\tan\b=40$ and 
$\tan\b=60$, and also show, for comparison, lower limits on 
the Higgs-boson mass obtained by direct searches at the CERN 
$e^+e^-$ collider LEP \cite{LEP:Working:Group} 
(vertical line at $m_{H^{\pm}}=78.5\ \GeV$) and 
by the measured $b\to s\g$ rate, using the leading-order result for the 
branching fraction.\label{rs:2hdm}}
\end{center}
\end{figure}

Using the constraints of the preceding sections, together 
with the results for the Wilson coefficients of scalar and pseudoscalar 
operators [\eq{constrained:2HDM:res}], we obtain exclusions 
in the $(m_{H^{\pm}}, R_S)$ plane, shown in \fig{rs:2hdm}.
%
% FIGURE 6
%
The observation to be noted here concerns the measured $b\to s\g$ branching 
fraction, which gives the 
strongest constraint on the Higgs-boson mass, thereby placing an upper bound
on $|R_S|$ of $0.17$.~As a consequence, the FB asymmetry due to non-standard 
scalar interactions turns out to be exceedingly small, 
typically of the order of 
$O(10^{-3})$. It should also be remarked 
that the lower bound on the charged Higgs-boson mass, 
$m_{H^{\pm}}\approx 260\ \GeV$, obtained from the measured inclusive 
$b\to s\g$ fraction in the context of the type-II 2HDM, 
does not directly apply to 
supersymmetric extensions of the SM since the chargino amplitude may 
interfere destructively with the charged Higgs- and $W$-boson contributions. 
Given a charged Higgs-boson mass of $m_{H^{\pm}}=260\ \GeV$, 
 the branching ratio of 
$\B_s\to\m^+\m^-$ amounts to $(1.4$--$4.8)\times 10^{-9}$ for 
$40\leqslant \tan\b\leqslant 60$, which should be compared with the
 SM result of
$(3.1\pm 1.4)\times 10^{-9}$ [\eq{SM:pred:BRbmumu}]. 
The average FB asymmetry is estimated 
to lie in the interval  $-0.7\times 10^{-3} \leqslant \av{A_{\text{FB}}}\leqslant
-1.6\times 10^{-3}$, which is much too small to be detected. 
For the decays $\B\to K \m^+\m^-$ and $\B\to K^*\m^+\m^-$,
we predict branching ratios of $5.2\times 10^{-7}$ and 
$1.7\times 10^{-6}$ respectively (to be compared with the SM expectations
of about $6 \times 10^{-7}$ and $2\times 10^{-6}$).
Note that these decays are 
largely unaffected by the 
charged Higgs-boson contributions to $R_9, R_{10}$, which are 
proportional to $\cot^2\b$, and hence small in the large $\tan\b$ regime.

Our conclusion is therefore that current experimental 
data on various rare $B$ decays -- apart from $b\to s\g$ -- do not provide any 
constraints on the parameter space in two-Higgs-doublet models of class II.   
Moreover, the predictions for the branching ratios of the $B$ decay modes 
under study are comparable to those of the SM.
We next turn to the SUSY scenario.

\subsubsection{SUSY with minimal flavour violation}
As mentioned earlier, we do not consider any $\cp$-violating effects, 
and consequently the SUSY parameters and mixing matrices discussed in 
the previous section can be taken to be real. We further assume the 
sneutrinos to be degenerate in mass so that $R_{\sneutrino}$, which enters 
the expression in \eq{susy:result:box}, reduces to the unit matrix. 

For the sake of illustration, we perform the numerical analysis for a light 
stop $\tilde{t}_1$, with large mixing  $\theta_{\tilde{t}}$,
and charginos with large Higgsino components. 
We impose the lower bounds 
on the SUSY particle masses as compiled by the Particle Data Group 
\cite{PDG}. In the case of a light scalar top quark, there are 
additional constraints coming from electroweak measurements
such as  the $\r$ parameter \cite{rho:parameter}. 
As for the constraint from $b\to s \g$, it is well known that within 
supersymmetry there are chargino-stop contributions, in addition to 
charged Higgs 
boson and $W$ boson loop contributions, which can  significantly  affect
the $b\to s\g$ decay rate in the large $\tan\b$ 
region, thereby leading to constraints on the parameter space.\footnote{For example, within the minimal supergravity model, most of the 
parameter space is ruled out for $\m<0$, where $\m$ denotes the Higgsino mass 
parameter,  while for $\m>0$, much of the 
parameter space is allowed by experimental data on $b\to s \g$
\cite{bertolini:etal,baer:etal}.} This is due to the fact that 
at large $\tan\b$,  the chargino loop contributions grow linearly with 
$\tan\b$ (see, e.g., \rf{largetanbeta:bsg}). Thus, in order to satisfy the 
$b\to s\g$ bounds, we must require that chargino and charged Higgs- and 
$W$-boson contributions interfere destructively, so that significant 
cancellations can occur. Note that in this case the sign of the Wilson 
coefficient $\cseff$ is opposite to the SM one.         
  
\begin{figure}
\begin{center}
\epsfig{file=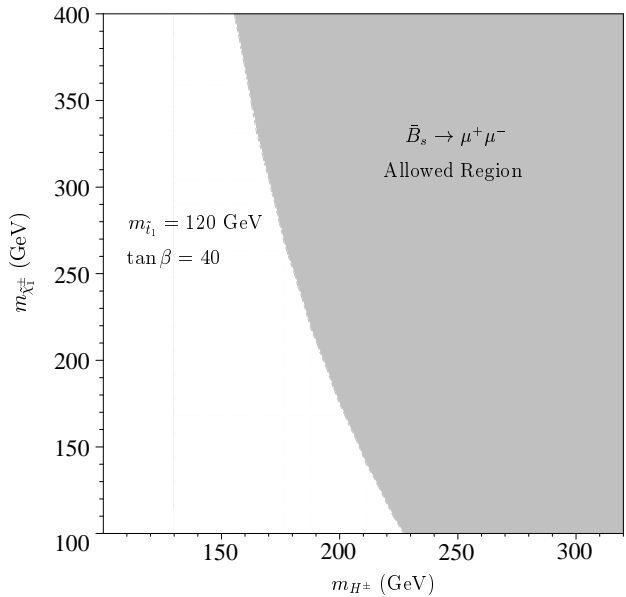,height=3.1in}\vspace{0.8em}
\caption{Allowed region in the $(m_{H^{\pm}}, m_{\chargino_1})$ plane 
derived from the upper bound on $\branch(\B_s \to \m^+\m^-$) for a 
light stop with $m_{\tilde{t}_1}=120\ \GeV$ and maximal mixing, 
$\theta_{\tilde{t}}\approx -45^\circ$.\label{susy:mH:mcharg1}}
\end{center}
\end{figure}

\begin{figure}
\begin{center}
\epsfig{file=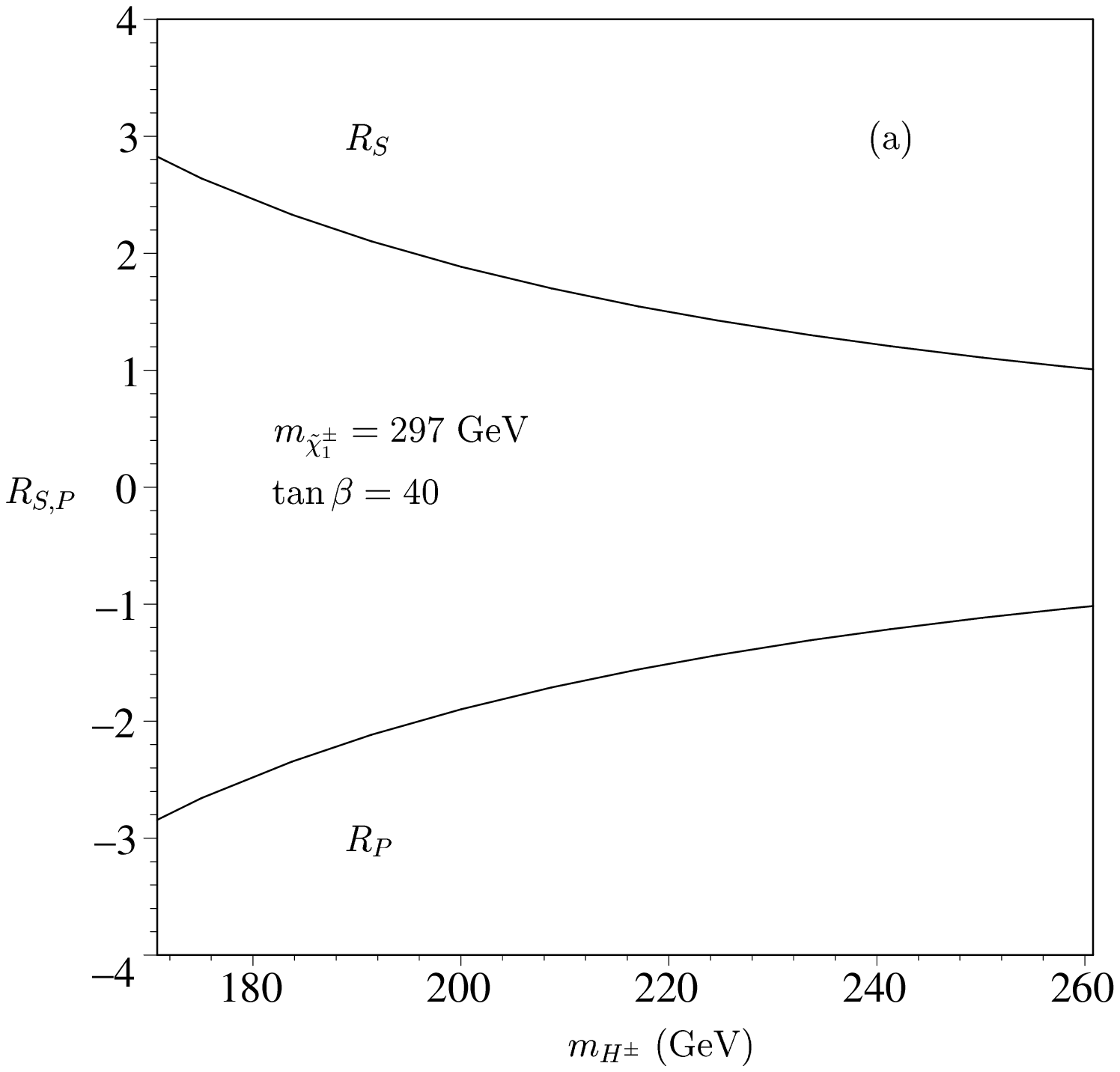,height=2.9in}\hspace{1em}
\epsfig{file=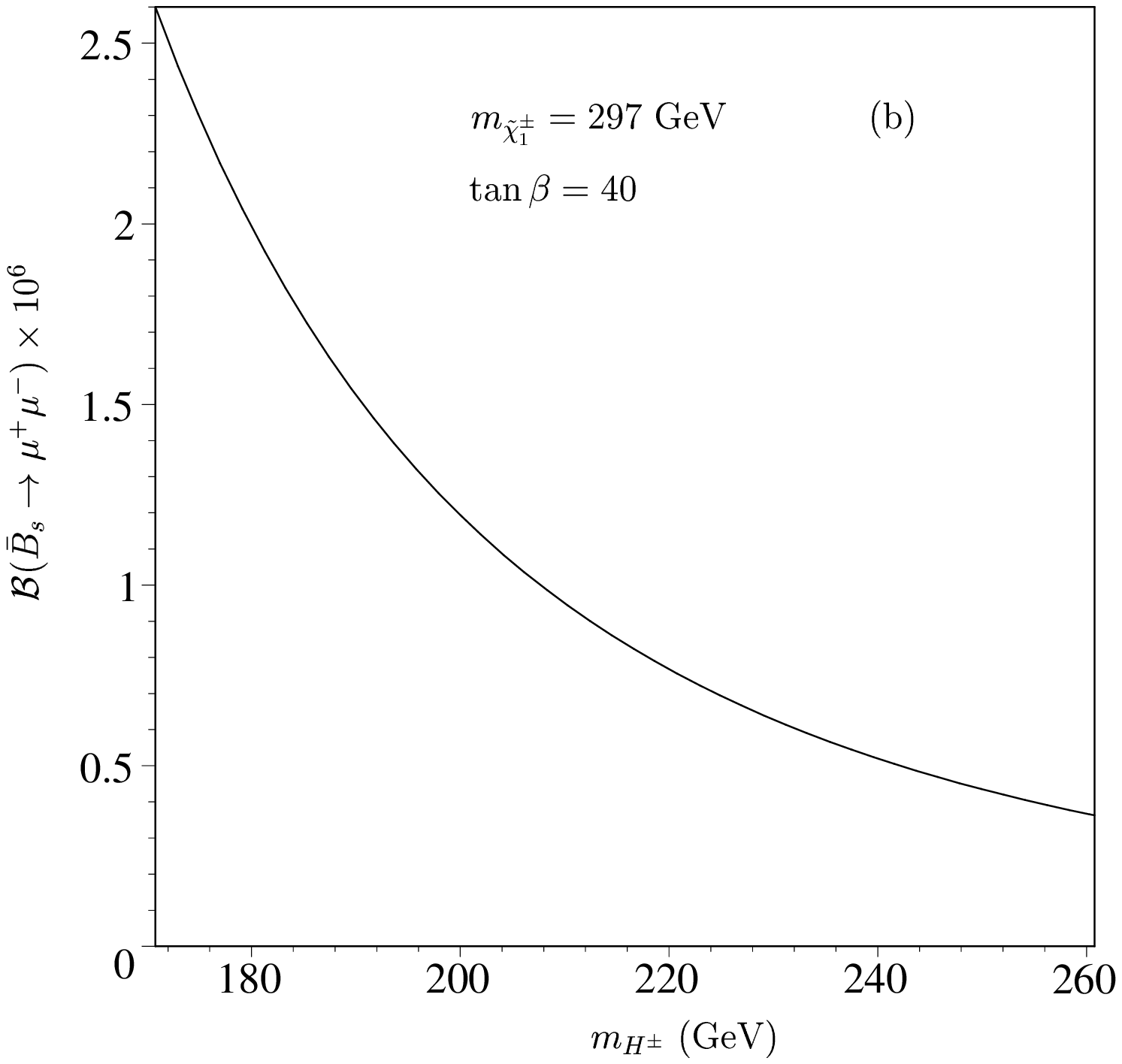,height=2.9in}\vspace{0.8em}
\caption{SUSY predictions for (a) $R_{S,P}$ and (b) the branching 
ratio of $\B_s\to \m^+\m^-$ at large $\tan\b$, as a function of the 
charged Higgs-boson mass, 
with $m_{\tilde{t}_1}=120\ \GeV$, $\theta_{\tilde{t}}\approx -45^\circ$,
and $m_{\chargino_1}=297\ \GeV$. We note that 
this choice of parameters satisfies all experimental bounds.\label{susy:mH:rsrp}}
\end{center}
\end{figure}

Using the expressions for the scalar and pseudoscalar Wilson coefficients,
Eqs.~(\ref{constrained:2HDM:res}), 
(\ref{susy:result:box})--(\ref{susy:result:count}), 
we obtain the allowed $(m_{H^{\pm}}, m_{\chargino_1})$  region
displayed in \fig{susy:mH:mcharg1} 
for $m_{\tilde{t}_1}=120\ \GeV$ and 
$\theta_{\tilde{t}}\approx -45^\circ$. As can be seen, the present upper limit
on $\branch(\B_s\to \m^+\m^-)$ already excludes a significant portion 
of the SUSY parameter
space with charged Higgs-boson and chargino masses. Remembering that the SM 
prediction for the branching ratio is of order $O(10^{-9})$, 
it is clear that within the context of SUSY, the 
$\B_s\to \m^+\m^-$ branching fraction can be increased
by several orders of magnitude, due to the $\tan^3\b$ enhancement of the 
counterterm diagrams [\eq{susy:result:count}].
%
% FIGURE 7
%
Given a chargino mass of $m_{\chargino_1}=297\ \GeV$,
the SUSY prediction  for $R_{S,P}$, consistent with the upper bound 
on $\branch(\B_s\to \m^+\m^-)$, is shown in \fig{susy:mH:rsrp}(a), 
as a function of 
the charged Higgs-boson mass. From this we infer  
that $R_{S,P}$ are  constrained to lie in the range 
$-3\lesssim R_{S,P}\lesssim3$, leading to an average 
FB asymmetry of less than $2\%$. Figure \ref{susy:mH:rsrp}(b) displays
the dependency of the $\B_s\to \m^+\m^-$ branching ratio on 
the charged Higgs-boson mass for $m_{\chargino_1}=297\ \GeV$.  
%
% FIGURE 8
%
Thus,
the measurement of the $\B_s\to \m^+\m^-$ branching 
fraction, together with the $b\to s \g$ bounds, provides a useful tool 
for constraining 
supersymmetric extensions of the SM. Finally, for the above parameter space 
point and $m_{H^{\pm}}=170\ \GeV$, we obtain $\branch(\B\to K \m^+\m^-)=1.0\times 10^{-6}$ and
$\branch(\B\to K^* \m^+\m^-)=2.6\times 10^{-6}$ to be compared with the present upper limits of $5.2\times 10^{-6}$ and $4.0\times 10^{-6}$. 

\section{Summary and conclusions}\label{discussion}
We have carried out a study of exclusive $B$ decays governed by the 
$b\to s l^+l^-$ transition in extensions of the SM with minimal 
flavour violation and new scalar and pseudoscalar interactions,
focusing on the dimuon final state, and  taking account of existing  
experimental data on $b\to s\g$ as well as the upper limits on
$\B_s\to \m^+\m^-$ and $\B\to K^{(*)}\m^+\m^-$ decays. We have restricted 
the discussion to the interesting case of large $\tan\b$, which may 
compensate for the inevitable suppression of scalar and pseudoscalar 
couplings by the lepton mass of $e$ or $\m$.  
Our main findings can be summarized as follows.

We have presented in a model-independent manner expressions for 
the  $\B_s\to l^+l^-$ branching fraction and the differential decay spectrum 
of $\B\to K l^+l^-$, together with the corresponding 
FB asymmetry, which is extremely tiny within the SM. 
In particular, we find that scalar and pseudoscalar interactions can, 
in principle, lead to striking effects in the decay distribution of 
$\B\to K\m^+\m^-$, while the branching ratio of
$\B\to K^*\m^+\m^-$ is essentially unaffected by the Higgs-boson 
contributions. We have demonstrated that once  
the constraint from $\B_s\to \m^+\m^-$ is taken into account, 
the effects of scalar and pseudoscalar couplings on the decay 
$\B\to K\m^+\m^-$ are much smaller. In view of the 
inherent uncertainty of the predictions for exclusive $B$ decays due to the form factors, 
it seems extremely unlikely that a 
measurement of the decay spectrum alone can  provide a clue to new 
physics with scalar and pseudoscalar interactions.  
We have also investigated the FB asymmetry of $\m^-$ in 
$\B\to K\m^+\m^-$ decay, which turns out to be at the level of a few per 
cent. Our analysis suggests that the observation of a nominal FB asymmetry 
of, say, $4\%$, together with a branching ratio of about 
$6\times 10^{-7}$,
will be challenging but might be feasible at the LHC and the Tevatron.   
As more precise data on the $\B_s\to \m^+\m^-$ branching ratio are 
available, more stringent upper limits will be placed on the FB asymmetry due 
to scalar interactions.
The essential conclusion of our model-independent analysis is that current 
experimental data on 
$\B_s\to \m^+\m^-$ decay already exclude
large values of the Wilson coefficients $c_S$ and $c_P$ of scalar and 
pseudoscalar operators, so that striking effects are not likely to  
show up in  the decay spectrum of $\B\to K\m^+\m^-$ and the 
corresponding FB asymmetry of the muon.
Even so, the FB asymmetry provides an independent
window to physics beyond the SM, especially to models with an extended 
Higgs sector, and its observation  would be an unambiguous signal of
new physics.

In extensions of the SM with minimal 
flavour violation, we have calculated the Higgs-boson contributions to the 
Wilson coefficients of scalar and pseudoscalar operators, and investigated how the new-physics parameters are 
constrained by existing experimental data on rare $B$ decays. 
Within the type-II 2HDM framework, 
where the Higgs sector corresponds to the one of 
the MSSM, we found no appreciable FB asymmetry or any large deviation from 
the SM prediction for the $\B\to K^{(*)}\m^+\m^-$ branching fractions. 
As for the decay 
$\B_s\to \m^+\m^-$, the branching ratio turns out to be in the range  
$(1.4$--$4.8)\times 10^{-9}$ for $40\leqslant \tan\b\leqslant 60$
and $m_{H^{\pm}}=260\ \GeV$, which is within the 
errors of the SM prediction of $(3.1\pm 1.4)\times 10^{-9}$. 
Ultimately, the smallness of the new-physics contributions
is caused by the mass of the charged Higgs boson, 
which is strongly constrained by the 
measured $b\to s \g$ branching fraction ($m_{H^{\pm}}\gtrsim 260\ \GeV$). 
We therefore conclude that within the type-II 2HDM there are no 
sizable new-physics effects on the $B$ decay modes described above, 
apart from $b\to s\g$.
By contrast, within SUSY, the effects of chargino and 
neutral Higgs-boson contributions on the $\B_s\to \m^+\m^-$ branching 
fraction can be enormous while satisfying the $b\to s \g$ bounds. 
We have considered a SUSY scenario 
with a scalar top quark with large mixing and a mass much 
lighter than the scalar partners of the light quarks. We find that 
for a given set of parameters obeying the $b\to s \g $ constraint,
the upper limit on $\branch(\B_s\to \m^+\m^-)$ severely constrains 
the masses of charginos and charged Higgs bosons. As a typical result, 
the lower bound $m_{H^{\pm}}\gtrsim 170\ \GeV$ has been derived for 
$m_{\chargino_1}=297\ \GeV$, $m_{\tilde{t}_1}=120\ \GeV$, and 
$\theta_{\tilde{t}}\approx -45^\circ$. 
The remaining observables are estimated to be $\av{A_{\text{FB}}}=1.8 \%$
and $\branch(\B\to K^{(*)} \m^+\m^-)=1.0\ (2.6)\times 10^{-6}$ 
(for $m_{H^{\pm}}=170\ \GeV$), the latter 
being close to the present upper limits. 
Clearly, the analysis of the decay 
$\B_s\to \m^+\m^-$ is complementary to the study of  
$\B\to K^{(*)}\m^+\m^-$ and $b\to s \g$ decays, which leads to constraints 
on the remaining short-distance coefficients, $\cseff, \ceff$, and $\cten$. 
A combined analysis of these 
decay modes, therefore, provides a powerful tool to constrain physics 
transcending the SM. 

\emph{Note added in proof.}
As this paper was readied for publication, we received a paper
by Huang \ea\ \cite{huang:etal:Erratum} 
that corrects the result presented for the 2HDM in 
\rf{huang:etal:recent}. The result of \rf{huang:etal:Erratum}
coincides with that of the present paper [\eq{2hdm:general:cs}].

\acknowledgments
We would like to thank Gerhard Buchalla, Andrzej J. Buras, and Manuel Drees
for useful discussions.~One of us (F.K.) would also like to thank Gudrun 
Hiller for communications.
We are indebted to Andrzej J. Buras for his comments on the manuscript.
This work was supported in part by the German `Bundesministerium f\"ur 
Bildung und Forschung' under contract 05HT9WOA0 and by the 
`Deutsche Forschungsgemeinschaft' (DFG) under grant Bu.706/1-1.
%
% Appendix 
%
\begin{appendix}\label{app:wilson:coeffs}
\section{Useful  functions}\label{aux:funcs}
\subsection{The one-loop function \bm$Y(s)$}
The function $Y(s)$ appearing in \eq{coeffs:sm} is given by \cite{eff:ham:sm}
\bea\label{ys:formula}
Y(s)&=& g(m_c,s)(3 c_1 + c_ 2 +3c_3 + c_4 + 3c_5 + c_6)
-\frac{1}{2} g(0,s)(c_3 + 3 c_4)\nnu\\
&-&\frac{1}{2}g(m_b,s)(4c_3 + 4c_4 +3c_5+c_6)
+\frac{2}{9}(3c_3 +c_4+3c_5+c_6),
\eea
where (at $\m_b=m_b^{\text{pole}}$)
\bea\label{loopfunc}
\lefteqn{g(m_i,s)=-\frac{8}{9}\ln(m_i/m_b^{\text{pole}})+\frac{8}{27}+\frac{4}{9}y_i
-\frac{2}{9}(2+y_i)\sqrt{|1-y_i|}}\nnu\\[.7ex]
&&\times\left\{
\Theta(1-y_i)\left[\ln\left(\frac{1 + \sqrt{1-y_i}}{1 - \sqrt{1-y_i}}\right)-i\p\right]+ \Theta(y_i-1) 2\arctan\frac{1}{\sqrt{y_i-1}}\right\},
\eea
with $y_i = 4m_i^2/s$, and 
\be
c_1=-0.249,\ c_2= 1.107,\ c_3=0.011,\ c_4=-0.025,\ c_5=0.007,\ c_6=-0.031.
\ee
In \eq{ys:formula}, we have omitted the one-gluon correction to the 
matrix element of the operator $\Oi_9$, which can be regarded as a 
contribution to the form factors.

\subsection{Auxiliary functions}
\be
B(x,y)= \frac{y}{x-y}\left[\frac{\ln y}{y-1}-\frac{\ln x}{x-1}\right],
\ee
\be
C(x,y)= \frac{y}{x-y}\left[\frac{x\ln x}{x-1}-\frac{y\ln y}{y-1}\right],
\ee
\be
D_1(x,y,z)= \frac{x\ln x}{(1-x)(x-y)(x-z)}
+ (x\leftrightarrow y) + (x\leftrightarrow z),
\ee
\be
D_2(x,y)= \frac{x\ln x}{(1-x)(x-y)}+(x\leftrightarrow y),
\ee
\be
D_3(x)= \frac{x\ln x}{1-x}.
\ee

\section{SUSY mass and mixing matrices}\label{app:susy:mass-matrices}
For the reader's convenience and to fix our notation, we give 
here the relevant mass and mixing 
matrices in the context of SUSY, for which we adopt the conventions of 
\rf{frank:jorge}.
\subsection{Chargino mass matrix}
Neglecting new $\cp$-violating phases, the chargino mass matrix is given by
\be\label{chargino:mass-matrix}
M_{\chargino}=
\left(
\begin{array}{cc}
M_2 & \sqrt{2} M_W \sin\b \\ 
\sqrt{2}M_W \cos\b & \m
\end{array}
\right),
\ee
where $M_2$ and $\m$ are the $W$-ino and Higgsino mass parameters 
respectively. 
This matrix can be cast in diagonal form by means of a biorthogonal
transformation
\be
U M_{\chargino}V^{\text{T}}=\diag(m_{\tilde{\chi}^{\pm}_1}, m_{\tilde{\chi}^{\pm}_2}),
\ee
$m_{\tilde{\chi}^{\pm}_{1,2}}$ being the chargino masses with 
$m^2_{\tilde{\chi}^{\pm}_1}<m^2_{\tilde{\chi}^{\pm}_2}$.
The orthogonal matrices $U$, $V$ read
\be
U=
\left(
\begin{array}{cc}
\cos\theta_U & \sin\theta_U \\ 
-\sin\theta_U&\cos\theta_U 
\end{array}
\right),\quad 
V=
\left(
\begin{array}{cc}
\cos\theta_V& \sin\theta_V \\ 
-\sin\theta_V &\cos\theta_V
\end{array}
\right),
\ee
with the mixing angles
\be
\sin 2\theta_{U,V}= \frac{2M_W\big[M_2^2+ \m^2 \pm (M_2^2-\m^2)\cos 2\b + 2\m
M_2\sin2\beta]^{1/2}}{m^2_{\tilde{\chi}^{\pm}_1}-m^2_{\tilde{\chi}^{\pm}_2}},
\ee
\be
\cos2\theta_{U,V}= \frac{M_2^2- \m^2\mp 2M_W^2 \cos2\beta}{m^2_{\tilde{\chi}^{\pm}_1}-m^2_{\tilde{\chi}^{\pm}_2}}.
\ee

\subsection{Scalar top mass matrix}  
In the $(\tilde{U}_L,\tilde{U}_R)$ basis, the $6\times 6$ up-type squark 
mass-squared matrix is given by 
\be\label{squark:massmatrix}
M^2_{\tilde{U}}=
\left(\begin{array}{cc}
M^2_{\tilde{U}_{LL}}&M^2_{\tilde{U}_{LR}}\\
M^2_{\tilde{U}_{LR}}& M^2_{\tilde{U}_{RR}}
\end{array}\right),  
\ee
which can be diagonalized by an orthogonal matrix $R_{\tilde{U}}$ such that 
\be
R_{\tilde{U}}M_{\tilde{U}}^2 R_{\tilde{U}}^{\text{T}}= 
\diag(m_{\tilde{u}_1}^2, m_{\tilde{u}_2}^2, \dots,m_{\tilde{u}_6}^2). 
\ee

Since we ignore flavour-mixing effects among squarks,
the matrix in \eq{squark:massmatrix} decomposes into a 
series of $2\times 2 $ matrices. Working in the so-called 
super-CKM basis \cite{super-ckm}, 
in which the quark mass matrices are diagonal, and 
squarks as well as quarks are rotated simultaneously, the LR terms in 
\eq{squark:massmatrix} are proportional to $m_q$, $q=u,c,t$. Thus,
large mixing can occur only in the scalar top quark sector, leading  
to a mass eigenstate $m_{\tilde{t}_1}$ possibly much lighter than the 
remaining squarks.

Defining the $6\times 3$ matrices 
\be\label{def:stt}
(\G^{U_L})_{ai}=(R_{\tilde{U}})_{ai},\quad  
(\G^{U_R})_{ai}=(R_{\tilde{U}})_{a, i+3},
\ee
we obtain
\be
(\G^{U_L})^{\text{T}}=\left(\begin{array}{cccccc}
1&0&0&0&0&0\\
0&1&0&0&0&0\\
0&0& \cos\theta_{\tsquark}&0&0&-\sin\theta_{\tsquark}
\end{array}\right), \quad
(\G^{U_R})^{\text{T}}=\left(\begin{array}{cccccc}
0&0&0&1&0&0\\
0&0&0&0&1&0\\
0&0& \sin\theta_{\tsquark}&0&0&\cos\theta_{\tsquark}
\end{array}\right),  
\ee 
with the mixing angle $(-\p/2\leqslant \theta_{\tsquark} \leqslant \pi/2$)
\be
\sin 2\theta_{\tsquark}=\frac{2m_t (A_t-\m\cot\b)}{m_{\tsquark_1}^2-m_{\tsquark_2}^2}, \quad
\cos 2\theta_{\tsquark}=\frac{(m_{\tsquark_L}^2 -m_{\tsquark_R}^2)+ \frac{1}{6}M_Z^2\cos 2\b(3-8\sin^2\theta_W)}{m_{\tsquark_1}^2-m_{\tsquark_2}^2}.
\ee
Here $A_t$ is the trilinear coupling, $m_{\tsquark_{L,R}}$ are the soft 
SUSY-breaking scalar masses, and $m_{\tsquark_{1,2}}$ denote the stop 
masses with $m^2_{\tsquark_1}< m^2_{\tsquark_2}$. 

\subsection{Sneutrino mixing matrix}
The $3\times 3$ mixing matrix $R_{\sneutrino}$ appearing in 
\eq{susy:result:box} is defined via
\be
R_{\sneutrino} M_{\sneutrino}^2 R_{\sneutrino}^{\text{T}}= 
\diag(m_{\sneutrino_1}^2, m_{\sneutrino_2}^2,m_{\sneutrino_3}^2),
\ee
where $M_{\sneutrino}^2$ is the sneutrino mass-squared matrix 
(see, e.g., \rfs{jorge,super-ckm}).
\end{appendix}
\newpage
%
%%%%%%%%%%%%% BIBLIOGRAPHY %%%%%%%%%%%%%%%%%%%%%%%%%%
%

%
\end{document}